\documentclass[final,sort,numbers,compress,3p]{elsarticle}

\usepackage{amssymb}

\usepackage{amsmath}
\usepackage{mathtools}
\usepackage{nicefrac}
\usepackage{amsfonts}
\usepackage{microtype}
\usepackage{natbib}
\usepackage{url}
\usepackage{xfrac}
\usepackage{algorithm}
\usepackage{algpseudocode}
\usepackage{import}
\usepackage[group-separator={,}]{siunitx}
\usepackage{mathrsfs}

\usepackage{color}

\usepackage{subcaption}
\usepackage{graphicx}
\usepackage[colorinlistoftodos]{todonotes}
\usepackage{booktabs}

\definecolor{mygray}{gray}{0.5}
\definecolor{tudcyan}{RGB}{0,166,214}
\usepackage[colorlinks, linkcolor=tudcyan, citecolor=tudcyan, urlcolor=tudcyan]{hyperref}
\usepackage[capitalise]{cleveref}

\usepackage{lineno}

\newcommand{\bs}[1]{\boldsymbol{#1}}
\newcommand{\rf}{\beta}

\makeatletter
\def\ps@pprintTitle{\let\@oddhead\@empty
 \let\@evenhead\@empty
 \let\@oddfoot\@empty
 \let\@evenfoot\@empty
}
\makeatother

\begin{document}
\date{November 19, 2024}

\begin{frontmatter}

\title{Integration of Active Learning and MCMC Sampling for Efficient Bayesian Calibration of Mechanical Properties}
\author[tudelft_citg]{Leon Riccius}
\ead{l.f.riccius@tudelft.nl}
\affiliation[tudelft_citg]{organization={Delft University of Technology, Faculty of Civil Engineering and Geosciences},
             city={Delft},
             country={The Netherlands}}
\affiliation[tudelft_diam]{organization={Delft University of Technology, Delft Institute of Applied Mathematics},
    city={Delft},
    country={The Netherlands}}

\author[tudelft_citg]{Iuri B.C.M. Rocha}
\author[tudelft_diam]{Joris Bierkens}
\author[tudelft_diam]{Hanne Kekkonen}
\author[tudelft_citg]{Frans P. van der Meer}

\title{}

\begin{abstract}

Recent advancements in Markov chain Monte Carlo (MCMC) sampling and surrogate modelling have significantly enhanced the feasibility of Bayesian analysis across engineering fields.
However, the selection and integration of surrogate models and cutting-edge MCMC algorithms, often depend on ad-hoc decisions.
A systematic assessment of their combined influence on analytical accuracy and efficiency is notably lacking.
The present work offers a comprehensive comparative study, employing a scalable case study in computational mechanics focused on the inference of spatially varying material parameters, that sheds light on the impact of methodological choices for surrogate modelling and sampling.
We show that a priori training of the surrogate model introduces large errors in the posterior estimation even in low to moderate dimensions.
We introduce a simple active learning strategy based on the path of the MCMC algorithm that is superior to all a priori trained models, and determine its training data requirements.
We demonstrate that the choice of the MCMC algorithm has only a small influence on the amount of training data but no significant influence on the accuracy of the resulting surrogate model.
Further, we show that the accuracy of the posterior estimation largely depends on the surrogate model, but not even a tailored surrogate guarantees convergence of the MCMC.
Finally, we identify the forward model as the bottleneck in the inference process, not the MCMC algorithm.
While related works focus on employing advanced MCMC algorithms, we demonstrate that the training data requirements render the surrogate modelling approach infeasible before the benefits of these gradient-based MCMC algorithms on cheap models can be reaped.

\end{abstract}


\end{frontmatter}

\section{Introduction}
\label{sec:intro}

\noindent
In engineering mechanics, complex computational models, e.g. with the finite element method (FEM), are used to simulate material behaviour.
The number of parameters for these models can be large, for instance if the material properties vary in space.
These parameters often represent an internal state of the model that is difficult or impossible to measure directly.
Therefore, an inverse problem must be solved to estimate these model parameters from indirect experimental measurements, commonly referred to as observations.
In the nonelastic regime, these inverse problems are typically non-linear, non-convex, and ill-posed, but they can be regularised elegantly by adopting a Bayesian approach.
This approach to inverse modelling employs Bayes' theorem to update the distribution of the parameters based on the observed data \cite{Dobrilla2023,Marsili2017}.
On the downside, Bayesian inverse modelling is computationally expensive, preventing its widespread application to large-scale problems.

Recent advancements in Markov chain Monte Carlo (MCMC) sampling and surrogate modelling have substantially improved the applicability of Bayesian analysis in these scenarios.
Despite the potential of these innovations, the selection of surrogate models and integration with state-of-the-art MCMC algorithms often rely on ad-hoc choices.
A systematic evaluation of their combined impact on overall performance remains scarce.
This lack of clarity obscures the understanding of how each component impacts the overall effectiveness of the inference process.
Addressing this gap, our study conducts a comparative analysis to elucidate the effects of the methodological decisions regarding surrogate modelling and sampling.

In Bayesian inference, a prior distribution for the parameters is assumed and combined with a likelihood function.
The likelihood measures the agreement of the model predictions with the observed system response.
Following the Bayesian formalism, a posterior distribution for the parameters is obtained.
For instance, when inferring material properties from measurements, the posterior distribution must be approximated due to the non-linear nature of the likelihood function, e.g. via sampling from the joint distribution of data and parameters.
The workflow of a Bayesian analysis entails the following steps:
1) The sampling algorithm generates a new state, i.e. a realisation of the parameter vector.
2) The forward model is evaluated on this state to produce model predictions.
3) The model predictions are compared to the observations to produce the likelihood.
4) The sampling algorithm determines whether the proposal is accepted or rejected, taking into account the prior, the likelihood, and the previous state.
The bottleneck of the sampling effort is the evaluation of the forward model (step 2), which, in case of FEM-based inverse modelling, is an expensive simulator that maps the parameters to the observables.
Steps 1 and 4 are implemented by the sampling algorithm and determine how often steps 2 and 3 are executed.
The most commonly used sampling algorithm is the random walk Metropolis (RWM) algorithm due to its simplicity and guaranteed asymptotic convergence~\cite{Mengersen1996}.
However, its sampling efficiency decreases with the dimensionality of the parameter space~\cite{Roberts2001}.

One way to reduce the number of costly forward passes is by employing sampling algorithms that more efficiently produce uncorrelated samples from the posterior distribution.
The first-order Metropolis adjusted Langevin algorithm~\cite{Roberts2002}, Hamiltonian Monte Carlo~\cite{Duane1987}, or their second-order manifold variants~\cite{Girolami2011}, as well as piecewise deterministic Markov process based algorithms~\cite{Bierkens2019} maintain high sampling efficiency in high-dimensional settings.
Empirical studies of their performance on engineering problems~\cite{Girolami2011,Chong2017,Goodman2010} suggests their theoretical performance in terms of convergence rates on canonical distributions~\cite{Gelman1997,Roberts1998} translates well to practical settings.
However, these algorithms require the gradient of the log-joint, or even higher derivatives thereof, which amounts to the computation of the model sensitivities with respect to its parameters.
These quantities are rarely available for large-scale computational mechanics simulators, preventing a straightforward application of these algorithms.
The gradients can be obtained with a finite difference scheme, but the computation of a finite difference stencil often negates the performance gains of the more advanced algorithms.

Ensemble-based methods such as transitional (ensemble) Markov chain Monte Carlo (TMCMC, TEMCMC)~\cite{Ching2007, Lye2022}, or sequential Monte Carlo in the form of Bayesian subset simulation with structural liability methods \cite{Straub2015} show significant performance improvements over RWM in terms of sampling efficiency.
While these algorithms do not require a differentiable forward model, they require at least $\mathcal{O}(10^4)$ model evaluations to ensure convergent posterior statistics.
This would still render a Bayesian analysis infeasible, even when assuming only a few minutes of runtime per model evaluation.

The other increasingly common strategy to reduce the computational cost of the sampling effort is the application of surrogate models.
Surrogate models exploit the redundancy in the model predictions required for collecting posterior samples.
Instead of running the expensive simulator in each MCMC step, a cheap-to-evaluate surrogate model is trained on a small number of solver calls.

Learning a cheap mapping from the parameter domain to the observations (step 2) is one of these surrogate modelling approaches.
\citet{YuanHu2024} replace a finite difference solver with a polynomial chaos surrogate model for the inference of thermal diffusivity coefficients.
\citet{Thomas2022} replace a FEM solver with a Gaussian process (GP) model to infer the orientation of microscopic fibers from macroscopic stress measurements in the elastic regime.
In~\citet{Wu2020a}, a neural network mapping from material parameters to strain paths takes the place of an iterative mean-field homogenisation algorithm for material property inference in the presence of plastic deformations.
\citet{Deveney2023} employ the deep Galerkin method instead of the classic finite difference solver to infer spatio-temporal heat fluxes in rotating disc systems.
All these approaches have in common that multiple models must be trained if the dataset consists of observations originating from fundamentally different conditions.
The three distinct test scenarios in~\citet{Wu2020a} motivate the training of three neural networks to reduce the complexity of individual tasks.
Further, the works employing polynomial chaos and Gaussian process models do not consider correlations between the outputs and train one model per scalar observation.
While model training is typically not the computational bottleneck in the settings we consider here, tweaking the hyperparameters for the individual models can be cumbersome and time-consuming for the analyst.

In step 3 of the workflow, the observations are funneled into the likelihood function to produce a scalar measure for the agreements of observations and model predictions for a given parameter vector.
Therefore, it is appealing to use a surrogate for the likelihood function directly instead of for the observations, with the clear advantage that only one model must be trained.
Following this idea~\citet{Chen2023} use a neural network to accelerate the parameter identification of an elastoplastic CuCrZr alloy.
Both~\citet{Drovandi2018} and~\citet{DelVal2022} employ a GP to efficiently infer stochastic rate constants in a gene network and material parameters governing the catalytic recombination, respectively.
However, targeting the likelihood function also introduces certain complexities.
The likelihood incorporates all model non-linearity at once and might be more difficult of a target than the individual model predictions.
Moreover, the likelihood has its own parameters, e.g. the observation noise, that one might be uncertain about.
These parameters of the observation model can be inferred alongside the model parameters, but this increases the dimensionality of the input space for the surrogate model.
Despite these challenges, we prioritise the likelihood approach in our study.
This decision is driven by its broader applicability across various domains, as it allows for a more generalised framework for modelling in contrast to methods tailored to specific physical models.
It should be noted that multi-fidelity MCMC~\cite{Zhang2018} and non-linear dimensionality reduction techniques~\cite{Dasgupta2024} have shown promising results for Bayesian inverse modelling.
However, these methods come with their own set of modelling challenges and therefore remain beyond the scope of this work.

For any choice of surrogate model, training data selection is crucial for the accuracy of the approximation.
Most commonly, an a priori strategy is employed.
The training points are either sampled from the prior, generated by Latin hypercube sampling~\cite{YuanHu2024, Thomas2022, Wu2020a, Chen2023}, or spread out on a grid~\cite{DelVal2022}.
While the a priori approach might produce sufficiently accurate models in low stochastic dimensions, it is susceptible to the curse of dimensionality:
the probability of being within a certain proximity of a training data point reduces exponentially with the dimensionality of the parameter space.
The model accuracy is typically measured by the mean squared error (MSE) on a validation set.
This set is generated with the same strategy as the training data.
However, this MSE is not necessarily a good indicator of the quality of the surrogate model for a Bayesian analysis ---
the surrogate model is only required to be accurate in regions of high posterior density.
Active learning strategies, such as~\cite{Deveney2023, Kandasamy2015, Drovandi2018, Dinkel2024}, exploit this more localised requirement.
They probe the posterior distribution to collect more informative data points.
It should be noted that surrogate models are generally not only cheap to evaluate, but also cheap to differentiate, enabling the application of gradient-based sampling algorithms.

The aforementioned works which focus on surrogate models for Bayesian inference identify MCMC sampling as a bottleneck and employ a variety of algorithms:
RWM in~\cite{Wu2020a, DelVal2022}; HMC or its No U-Turn (NUTS) variant ~\cite{Deveney2023, YuanHu2024, Thomas2022}; TMCMC ~\cite{Chen2023}.
However, the specific choices are rarely justified or even commented on.
A good overview of the \textit{mix and match} regarding surrogate and sampling algorithm can be found in the review of \citet{Hou2021} on Bayesian inference for building energy models.
The review contains an assessment both of surrogate models and sampling algorithms in isolation, but does not cover the integration of both components into a single framework.
Especially for active learning strategies, the interplay between the surrogate model and the sampling algorithm is crucial:
the sampler determines where the surrogate model is queried (step 2), and the surrogate models in turn affects the acceptance probability (step 4), and the subsequent proposal generation (step 1).

In this work, we investigate the interplay of surrogate modelling and sampling algorithms for Bayesian inference.
We focus on the inference of material parameters in the non-linear regime of a computational mechanics model.
We introduce a scalable case study to create a series of inference problems with comparable complexity but increasing stochastic dimensionality.
We employ a GP surrogate model for the likelihood and propose to follow the MCMC chain to collect training data.
This form of active learning is based on uncertainty of the model prediction, informing our choice of the GP as surrogate model.
As we will demonstrate in~\cref{sec:online_vs_offline}, active learning is a crucial component of the framework, ruling out most of the other surrogate models.

Using this set of problems we investigate the following aspects of surrogate accelerated Bayesian inference:
i) How does the choice of data collection strategy impact the accuracy of the surrogate model?
ii) Do superior MCMC algorithms construct better surrogate models?
iii) After training the surrogate model, how does the sampling algorithm for the generation of posterior samples impact accuracy and performance?

\section{Background}\label{sec:background}
\subsection{Computational mechanics}\label{sec:computational-mechanics}

\noindent
This section presents a brief introduction to computational mechanics.
The deformation of the body within the domain $\mathscr{D}$ with boundary $\mathscr{B}$ is governed by the equilibrium equation, which can be written as

\begin{align}
    \nabla \cdot \bs{\sigma}  &= \bs{0} &&\mathrm{for} \, \bs{x} \in \mathscr{D} \label{eq:equilibrium} \\
    \bs{u}(\bs{x},t) &= \hat{\bs{u}}(\bs{x},t) &&\mathrm{for} \, \bs{x} \in \mathscr{B} \label{eq:dirichlet_bc},
\end{align}

\noindent
where $\bs{\sigma}$ is the stress tensor and inertia effects and external forces are assumed to be negligible.
The boundary conditions in \cref{eq:dirichlet_bc} enforce a displacement $\hat{\bs{u}}$ at the boundaries of the domain.
The kinematics of the deformation are described by the displacement field $\bs{u}(\bs{x})$.
Assuming small strains, the strain tensor $\bs{\varepsilon}$ is given by

\begin{align}
    \bs{\varepsilon} &= \frac{1}{2} \left( \nabla \bs{u} + \left(\nabla \bs{u} \right)^\top \right). \label{eq:kinematic_relation}
\end{align}

\noindent
This kinematic relation must be linked to the equilibrium equation in \cref{eq:equilibrium}.
The constitutive model

\begin{align}
    \bs{\sigma} = \mathcal{C}(\bs{\varepsilon}, \rf), \label{eq:constitutive_model}
\end{align}

\noindent
relates the stresses to the strains and a local material parameter $\beta$.
It should be noted that $\beta$ does not represent an evolving internal state as history dependent materials are not considered in this work.
In many situations, the local parameter value depends on a set of global parameters $\bs{\theta} \in \mathbb{R}^d$ and the location $\bs{x}$, i.e. $\rf = \rf(\bs{x}, \bs{\theta})$.
Substituting \cref{eq:constitutive_model} into \cref{eq:equilibrium} gives

\begin{align}
    \nabla \cdot \left(\mathcal{C}\left(\bs{\varepsilon}, \rf \right) \right) = \bs{0} \label{eq:balance_of_linear_momentum_2},
\end{align}

\noindent
which together with \cref{eq:kinematic_relation} can be solved for the displacement field $\bs{u}(\bs{x})$ and yields the forward model $\mathcal{M}(\bs{\theta})$.
The material model may depend on multiple parameters that could also exhibit spatial variability.
While the extension to the multivariate case is straightforward, we will focus on the scalar case here for simplicity.

The choice of constitutive model—be it linear or non-linear, isotropic or anisotropic, and dependent or independent of time—hinges on the specific material and its regime of deformation.
Irrespective of the clarity in the nature of the chosen constitutive model, a persistent challenge lies in the inherent uncertainty of material parameters \cite{Rappel2019,Rappel2020}.
These parameters, crucial for accurately predicting material behaviour, are often not directly measurable but must be inferred from experimental data.
The task of inferring these parameters frames the inverse problem associated with the forward model~$\mathcal{M}$. \subsection{Random Fields and their Discretisation}\label{sec:random-fields}

\noindent
Acknowledging the lack of total control over manufacturing processes or the inherent randomness of the material itself, the assumption of homogeneous material properties is generally not justified.
In many applications, material properties can vary significantly across different regions of a part and exhibit randomness under repeated production.
We are, therefore, interested not just in inferring global material parameters but also how they vary throughout $\mathscr{D}$.
However, there is typically some structure to the randomness that we need to consider when inferring the spatial distribution of material properties.

A useful way to capture this spatial variability is through random fields, which provide a framework for describing quantities that vary both in space and with some inherent randomness.
More specifically, a random field is a collection of random variables associated with points in a physical space, $x \in \mathscr{D} \subset \mathbb{R}^D$.
For simplicity, we can think of a random field as a function $\hat{\rf}(\bs{x})$ that assigns random values to each point $\bs{x}$ in the domain.

Many physical quantities, such as material properties, can be modeled as Gaussian random fields.
This means that any set of values the field takes at different points, $\hat{\rf}(\bs{x}_1), \dots, \hat{\rf}(\bs{x}_k)$ for $\bs{x}_1, \dots \bs{x}_k \in \mathscr{D}$, follows a multivariate normal distribution.
This type of field is fully described by two key statistical properties:
i) the mean function $\mu_{\hat{\rf}}(\bs{x})$, which describes the average value of the field at each point, and
ii) the covariance function $C_{\hat{\rf}}(\bs{x}, \bs{x}')$, which tells us how values at different points are correlated based on their distance apart.

In practice, the field is expressed as a truncated series expansion in terms of a set of deterministic basis functions $\phi_i(\bs{x})$ with random coefficients $\theta_i$.
By cutting off the series expansion at a finite number of terms $d$, we obtain a finite-dimensional approximation of the random field:

\begin{align}
    \label{eq:random_field_expansion}
    \hat{\rf}(\bs{x}) \approx \rf(\bs{x}, \bs{\theta}) = \mu_{\rf}(\bs{x}) + \sum_{i=1}^{d} \phi_i(\bs{x}) \theta_i.
\end{align}

\noindent The expansion effectively amounts to a separation of the physical and the stochastic domain~\cite{Sudret2000}.
Since all randomness is now contained in the coefficients $\bs{\theta} \in \mathbb{R}^d$, inferring the random field reduces to inferring $\bs{\theta}$.

Expansions encountered in related works employ a basis of polynomials~\cite{Deveney2023,Marzouk2009}, wavelets~\cite{Nouy2009}, B-splines~\cite{Vigliotti2018}, or Gaussian-process shape functions~\cite{Li1993}.
The optimal expansion in terms of the total mean squared error of the truncation compared to the full sum is the Karhunen-Loève (KL) expansion.
It is the eigenfunction expansion of the covariance function of the random field~\cite{Ghanem1991}.
For a Gaussian random field, the KL expansion coefficients $\bs{\theta}$ are i.i.d. standard normal random variables.
For the other expansions, the coefficients are normally distributed but correlated.

We opt for a radial basis function (RBF) expansion.
The basis is given by

\begin{align}
    \phi_i(\bs{x}) = \exp\left(-\frac{1}{2l^2} \|\bs{x} - \bs{c}_{i}\|^2\right),
\end{align}

\noindent
where $\bs{c}_i$ are the centers of the RBFs and $l$ is the length-scale.
Each basis function $\phi_i$ reaches its maximum at $\bs{c}_i$ and decays with increasing distance from the center.
This choice is not optimal in terms of the total mean squared error for a given number of basis functions.
However, it has a clear advantage for defining a sequence of comparable inference problems:
the RBF expansion leads to a parameterisation of the field where all coefficients have a similar influence on the global system response.
On the other hand, the ordering of the eigenfunctions of the KL expansion according to their eigenvalues ensures that the model is most sensitive to changes in the first few coefficients.
Furthermore, we circumvent the need to solve the associated eigenvalue problem.

Given that the random field is characterised by a squared exponential covariance function, employing a linear model with RBFs and appropriately setting the RBF length-scale produces a similar covariance structure.
It converges to the desired covariance in the limit of infinitely many basis functions~\cite[p. 84]{Rasmussen2006}.
Note that an RBF model with length-scale $l$ produces a covariance function with effective length-scale $\ell = \sqrt{2} l$~\cite{Mackay1998}.
For simplicity, we choose the prior of the weights to be i.i.d Gaussian, i.e. $\bs{\theta} \sim \mathcal{N}(\bs{0}, \sigma_\theta^2 \bs{I})$.
It is convenient to store the basis functions in a feature vector $\bs{\phi}(\bs{x}) = [\phi_1(\bs{x}), \ldots, \phi_{d}(\bs{x})]^\mathsf{T}$.
The prior mean of the field then reads

\begin{align}
    \mathbb{E}[\rf(\bs{x}, \bs{\theta})]
    = \mathbb{E}\left[\mu_{\rf}(\bs{x}) + \bs{\phi}(\bs{x})^\mathsf{T} \bs{\theta}\right]
    = \mu_{\rf}(\bs{x}) + \bs{\phi}(\bs{x})^\mathsf{T} \mathbb{E}[\bs{\theta}]
    = \mu_{\rf}(\bs{x}),
\end{align}

\noindent
showing the desired mean can be achieved by setting $\mu_{\beta} = \mu_{\hat{\beta}}$.
The field's covariance function is given by

\begin{align}
    \mathrm{cov}(\rf(\bs{x}, \bs{\theta}), \rf(\bs{x}', \bs{\theta}))
    &= \mathbb{E}[(\rf(\bs{x}, \bs{\theta}) - \mu_{\rf}(\bs{x}))(\rf(\bs{x}', \bs{\theta}) - \mu_{\rf}(\bs{x}'))]\\
    &= \mathbb{E}[\bs{\phi}(\bs{x})^\mathsf{T} \bs{\theta} \bs{\theta}^\mathsf{T} \bs{\phi}(\bs{x}')] = \bs{\phi}(\bs{x})^\mathsf{T} \bs{\phi}(\bs{x}') \sigma_\theta^2.
\end{align}

\noindent
The variance consequently is

\begin{align}
    \sigma^2_{\rf}(\bs{x}) = \bs{\phi}(\bs{x})^\mathsf{T} \bs{\phi}(\bs{x}) \sigma_\theta^2.
\end{align}

\noindent The prior variance of the weights $\sigma_\theta^2$ can be set to match the variance of the random field of interest.
It is then given by

\begin{align}
    \sigma_\theta^2 = \frac{\sigma_{\hat{\rf}}^2}{\sup \left\{\bs{\phi}(\bs{x})^\mathsf{T} \bs{\phi}(\bs{x}), \bs{x} \in \mathscr{D} \right\}}. \label{eq:rbf_variance}
\end{align}

\noindent
The random field parameters, i.e. its mean $\mu_{\hat{\rf}}$, variance $\sigma_{\hat{\rf}}^2$, and effective length-scale $\ell$, can be inferred along the coefficients $\bs{\theta}$ in a hierarchical Bayesian model.
We assume to have prior knowledge of the problem that we can use to set these hyperparameters, as hierarchical modelling is not the focus of this study.

\begin{figure}[h]
    \centering
    \includegraphics[width=0.9\textwidth]{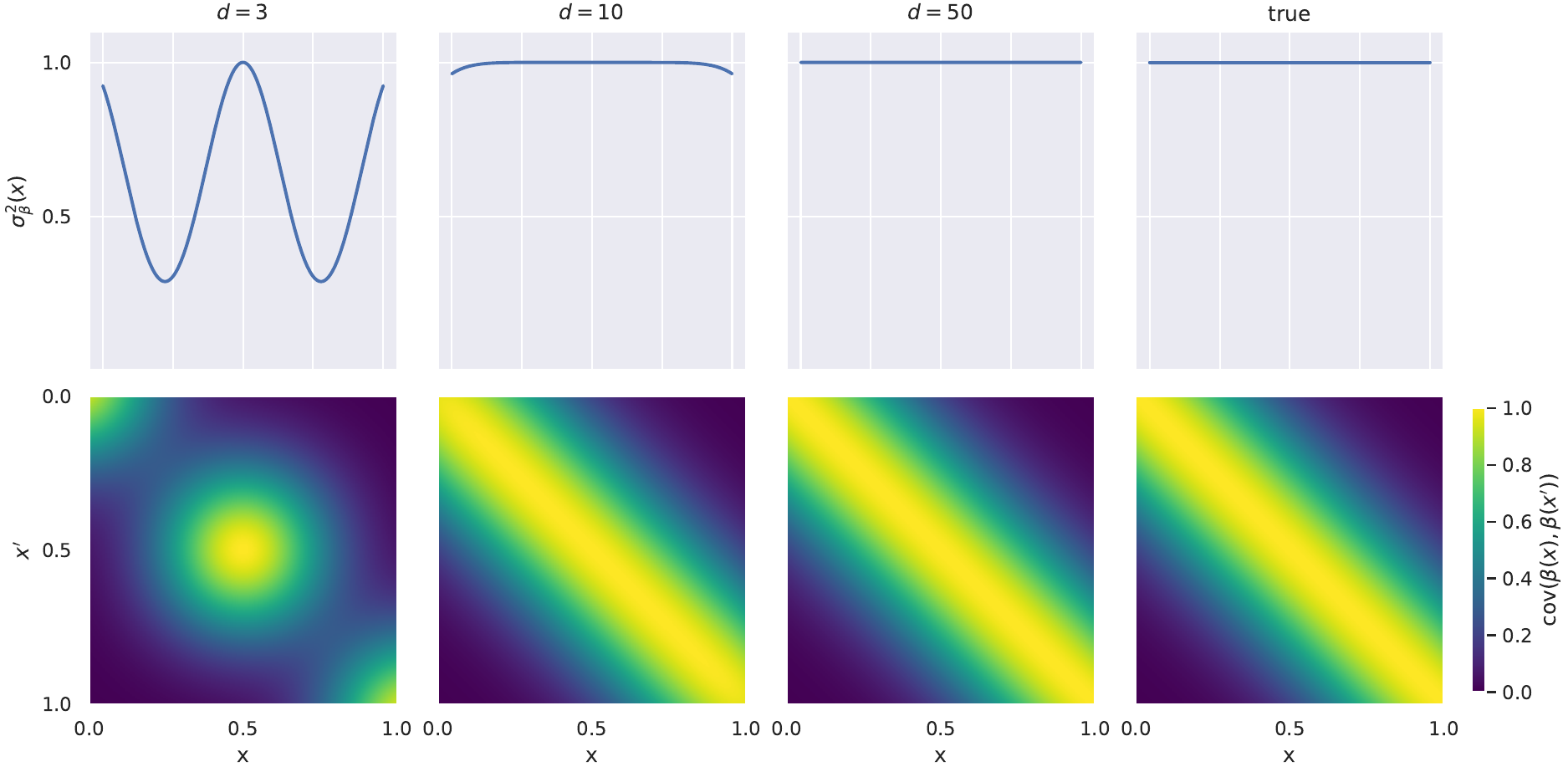}
    \caption{Discretisation of the random field with increasing amount of radial basis functions. The variance of the fields is shown in the top row. The banded covariance structure is displayed in the bottom row.}
    \label{fig:rbf_basis}
\end{figure}

\cref{fig:rbf_basis} shows the discretisation of a random field on a one-dimensional domain $\mathscr{D}$ with varying amounts of basis functions.
As the number of basis functions increases, the variance of the field approaches a constant value throughout the domain.
The banded covariance structure on the far right of the figure is well approximated by the RBF model.
We can see that the expansion produces a stationary random field that converges in variance and length-scale to the infinite-dimensional random field.

 \subsection{Bayesian Inference for Random Fields}\label{sec:bayesian-inference}

\noindent
Our PDE model $\mathcal{M}$ maps material parameters $\bs{\theta} \in \mathbb{R}^d$ to an output $\bs{y} \in \mathbb{R}^m$, i.e. $\bs{y} = \mathcal{M}(\bs{\theta})$.
The inverse problem is concerned with inferring these model parameters from observation data $\mathcal D = \tilde{\bs{y}}$.
In mechanics, the observation data $\mathcal D$ typically comprises noisy measurements of the displacement field $\bs{u}$ and the reaction forces $\bs{f}$, i.e. $\tilde{\bs{y}} = [\bs{u}^\mathsf{T}, \bs{f}^{\mathsf{T}}]^{\mathsf{T}}$.
For a Bayesian approach to inverse problems, the parameters and the data are represented as random vectors and equipped with a probability measure, see~\citet{Stuart2010} for a detailed description.
The prior distribution for the parameters $p(\bs{\theta})$ expresses all knowledge available on the parameters $\bs{\theta}$ prior to seeing any data.
The likelihood function $p(\mathcal{D}|\bs{\theta})$ expresses how well the model prediction matches the observed data for a given realisation of the parameter vector $\bs{\theta}$.
In our case, the likelihood is a function of the model $\mathcal{M}$, which allows the flow of information from the outputs $\bs{y}$ back to the model parameters $\bs{\theta}$.

Following Bayes theorem, the probability of the parameters $\bs{\theta}$ conditioned on the observations $\mathcal{D}$ reads

\begin{align}
    \underbrace{p(\bs{\theta}|\mathcal{D})}_{\mathrm{posterior}} =
    \frac{\overbrace{p(\mathcal{D} | \bs{\theta})}^{\mathrm{likelihood}}
        \overbrace{p( \bs{\theta})}^{\mathrm{prior}}}
    {\underbrace{p(\mathcal{D})}_{\mathrm{model \, evidence}}},
\end{align}

\noindent where the denominator on the right hand side is the so-called evidence.
This normalising constant can be obtained by marginalising out the parameters:

\begin{align}
    \label{eq:evidence}
    p(\mathcal{D}) = \int p(\mathcal{D} | \bs{\theta} ) p( \bs{\theta}) d \bs{\theta}.
\end{align}

\noindent
With a few exceptions, the integral in \cref{eq:evidence} is intractable.
While the evidence can be approximated via numerical methods such as Monte Carlo simulation, it is notoriously hard to compute accurately.
More efficiently, we can draw samples directly from the unnormalised posterior distribution via MCMC methods.
For the latter approach, we only need to know the posterior distribution up to a proportionality constant, hence no need to obtain the evidence.
We can draw samples from the posterior despite only having access to the joint distribution of data and parameters:

\begin{align}
    p(\bs{\theta} | \mathcal{D} ) \propto
    p(\mathcal{D}, \bs{\theta}) = p(\mathcal{D} | \bs{\theta}) p(\bs{\theta}).
    \label{eq:joint_distribution}
\end{align}

\noindent
The FEM solver only depends on the realisation of the random field $\hat{\rf}$ \cref{eq:random_field_expansion} at the integration points of the FEM mesh $\bs{\rf}_{\mathrm{f}}$.
The conditional distribution for observing the data is given by

\begin{align}
    p(\mathcal{D} | \rf ) = p(\mathcal{D} | \bs{\rf}_{\mathrm{f}})
\end{align}

\noindent
Hence, the joint posterior distribution for $\bs{\rf}_{\mathrm{f}}$ and $\bs{\theta}$ is

\begin{align}
    p(\bs{\rf}_{\mathrm{f}}, \bs{\theta} | \mathcal{D} ) \propto p(\mathcal{D} | \bs{\rf}_{\mathrm{f}}) p(\bs{\rf}_{\mathrm{f}} | \bs{\theta}) p(\bs{\theta}).
\end{align}

\noindent
The posterior distribution for $\bs{\theta}$ is obtained by marginalising out $\bs{\rf}_{\mathrm{f}}$

\begin{align}
    \label{eq:gp_mean_post}
    p(\bs{\theta} | \mathcal{D}) \propto \int p( \mathcal{D} | \bs{\rf}_{\mathrm{f}}) p(\bs{\rf}_{\mathrm{f}} | \bs{\theta}) p(\bs{\theta}) d \bs{\rf}_{\mathrm{f}},
\end{align}

\noindent
where all randomness is contained in the coefficients $\bs{\theta}$.
The values of the field in the FEM integration points and the rest of the domain depend on realisations of the coefficients deterministically, i.e. their distributions given $\bs{\theta}$ are Dirac measures:

\begin{align}
    p(\rf(x)|\bs{\theta}) &= \delta ( \mu_{\rf}(\bs{x}) + \bs{\phi}(\bs{x})^\mathsf{T} \bs{\theta} ), \label{eq:dirac_Fx}\\
    p(\bs{\rf}_{\mathrm{f}}|\bs{\theta}) &= \delta ( \bs{\mu}_{\rf_{\mathrm{f}}} + \bs{\Phi}_{\mathrm{f}}^\mathsf{T} \bs{\theta} ) \label{eq:dirac_Ff},
\end{align}

\noindent
where $\bs{\Phi}_{\mathrm{f}} = [\bs{\phi}(\bs{x}_1), \dots, \bs{\phi}(\bs{x}_{N_\mathrm{f}})] \in \mathbb{R}^{d \times N_{\mathrm{f}}}$ are the RBFs evaluated at the FEM integration points.
Making use of \cref{eq:dirac_Ff}, the integral in \cref{eq:gp_mean_post} evaluates to

\begin{align}
    p(\bs{\theta} | \mathcal{D}) \propto p( \mathcal{D} | \bs{\mu}_{\rf_{\mathrm{f}}} + \bs{\Phi}_\mathrm{f}^\mathsf{T} \bs{\theta} ) p(\bs{\theta}).
    \label{eq:post_field_params}
\end{align}

\noindent
The full field posterior is then simply obtained by plugging $p(\bs{\theta}|\mathcal{D})$ back into the expansion~\cref{eq:dirac_Fx}.

We have to define an observation model to complete the statistical model.
The dataset is an additional source of uncertainty as observations are typically corrupted by noise.
This noise can stem, e.g. from inherent randomness of the measuring device.
In many situations additive noise that is independent of the model parameters is assumed:

\begin{align}
    \label{eq:observation_model}
    \tilde{\bs{y}} = \mathcal{M} (\bs{\theta}) + \bs{\epsilon}
\end{align}

\noindent
where $\bs{\epsilon} \in \mathbb{R}^m$ is the noise random vector.
In this work, the noise vector is assumed to follow a zero-mean Gaussian distribution with covariance matrix $\bs{\Sigma}_{\epsilon \epsilon} \in \mathbb{R}^{m \times m}$.
More specifically, each data point comprises displacement and force observations $\bs{f}$ and $\bs{u}$, respectively, which come from measuring devices with distinct noise levels for $\bs{\epsilon}_u$ and $\bs{\epsilon}_f$.
Extending \cref{eq:observation_model} to the multi-output case, we have

\begin{align}
    \tilde{\bs{u}} = \bs{u} + \bs{\epsilon}_u, \quad \tilde{\bs{f}} = \bs{f} + \bs{\epsilon}_f.
\end{align}

\noindent
Assuming no correlation between the two types of observations, the likelihood takes the form

\begin{align}
    p \left( \begin{bmatrix} \bs{u} \\ \bs{f} \end{bmatrix} \bigg\rvert \, \bs{\theta} \right)
    = \mathcal{N} \left( \begin{bmatrix} \bs{u} \\ \bs{f} \end{bmatrix} \bigg\rvert \, \mathcal{M}(\bs{\theta}),
    \begin{bmatrix} \bs{\Sigma}_{uu} & \bs{0} \\
    \bs{0} & \bs{\Sigma}_{f\!f} \label{eq:likelihood}
    \end{bmatrix}\right)
\end{align}

\noindent
where the covariance structures $\bs{\Sigma}_{uu}$ and $\bs{\Sigma}_{f\!f}$ of the displacement and force observations, respectively, are known a priori.
It should be noted that the i.i.d.\ noise model choice is based on the assumption that we exactly know the physics of the problem, i.e. there is no model misspecification, and that our measurement devices show no spatial or temporal correlation.
This simplification is justified for the scope of this study, as it investigates aspects of the Bayesian inference workflow other than the definition of the likelihood function.
In practice, however, a more complex observation model that reflects all sources of uncertainty, as outlined in~\cite{Simoen2013}, should be employed.
Otherwise, the inference might yield overconfident posterior beliefs about the parameters governing the random field \cref{eq:post_field_params}, which can be misleading in a subsequent decision-making process.
 \subsection{Markov Chain Monte Carlo Methods}\label{sec:markov-chain-monte-carlo-methods}

\noindent
MCMC methods are a class of algorithms used to sample from a probability distribution from which direct sampling is difficult, hence they are suitable for Bayesian inference in computational mechanics.
A Markov chain that has the desired distribution as its equilibrium distribution is constructed and its states are recorded.
The more steps the chain takes, the more the distribution of the states of the chain converges to the target distribution.

\subsubsection{Random Walk Metropolis Algorithm}\label{subsec:metropolis-hastings-algorithm}

\noindent
The RWM algorithm is based on a random walk in the state space, determined by a proposal distribution to generate new states.
In the subsequent Metropolis correction, the proposed state is either accepted so that the chain moves, or rejected so that the chain stays in place.
This correction is essential for the convergence of the chain to the target distribution and is a core building block for many other MCMC algorithms.

Suppose we have a function $f(\bs{\theta})$ that is proportional to our target distribution $p(\bs{\theta}|\mathcal{D})$.
In our case, this function is the joint distribution of parameters and data $p(\mathcal{D}, \bs{\theta})$.
Once the data is observed, the joint distribution essentially becomes a function of the parameters $\bs{\theta}$ only.
Given the current state $\bs{\theta}^n$, a move to $\bs{\theta}^*$ is proposed according to the proposal distribution $q(\bs{\theta}^*|\bs{\theta}^{n})$.
Next, the Metropolis-Hastings ratio is calculated as

\begin{align}
    r(\bs{\theta}^n, \bs{\theta}^*) = \frac{f(\bs{\theta}^*)q(\bs{\theta}^*|\bs{\theta}^{n})}{f(\bs{\theta}^{n})q(\bs{\theta}^{n}|\bs{\theta}^*)}.
    \label{eq:metropolis-hastings_ratio}
\end{align}

\noindent
The move from $\bs{\theta}^{n}$ to $\bs{\theta}^*$ is then accepted with probability

\begin{align}
    a(\bs{\theta}^{n}, \bs{\theta}^*) = \min \left( 1, r(\bs{\theta}^{n}, \bs{\theta}^*) \right).
    \label{eq:acceptance_probability}
\end{align}

\noindent
If the proposal distribution $q(\bs{\theta}^*,\bs{\theta}^{n})$ is symmetric---the normal distribution is a common choice---the ratio simplifies to the Metropolis ratio

\begin{align}
    r(\bs{\theta}^{n},\bs{\theta}^*) = \frac{f(\bs{\theta}^*)}{f(\bs{\theta}^{n})},
    \label{eq:metropolis_ratio}
\end{align}

\noindent
also reflected in the algorithms name.
The Gaussian proposal mechanism can then be written as

\begin{align}
    \bs{\theta}^* = \bs{\theta}^n + s \sqrt{\bs{M}}\bs{z}^n,
    \label{eq:gaussian_proposal}
\end{align}

\noindent
where $\bs{M}$ is a positive semi-definite preconditioner matrix which defines the covariance of the proposal distribution, $\bs{z}^n$ is a vector of i.i.d. standard normal random variables, and $s$ is the proposal variance controlling the step size.
The corresponding proposal distribution then is

\begin{align}
    q(\bs{\theta}^*|\bs{\theta}^n) = \mathcal{N} (\bs{\theta}^*| \bs{\theta}^n, s^2 \bs{M}).
\end{align}

\noindent
The performance of the algorithm can be significantly improved by setting $\bs{M}$ to the sample covariance.
The sample covariance can either be estimated from a preliminary run of the algorithm or with a Laplace approximation of the posterior.
The proposal covariance can also be adapted continuously during runtime, as long as the level of adaptation is vanishing with the number of iterations~\cite{Roberts2007}.
Because diminishing adaptiveness is difficult to prove in practice, we set the proposal covariance to the sample covariance only once after the first 5000 MCMC-steps.
Until then, we use the prior covariance as preconditioner.
This version of the RWM algorithm serves as the baseline MCMC algorithm in our study.

High acceptance rates can be achieved by reducing the variance of the proposal distribution, but this can lead to slow exploration of the state space.
Conversely, a high variance can lead to fast exploration, but with a low acceptance rate.
It is, therefore, important to choose the step size $s$ in a way that balances these two aspects.
However,~\citet{Gelman1997} have shown that the proposal variance $s$ must be scaled with the state space dimensionality $d$ according to $\mathcal{O}(d^{-1})$ to achieve the optimal acceptance rate of $0.234$.
Hence, if $d$ is large, the transitions are small and the algorithm shows poor mixing and slow convergence.
Nonetheless, this poor scaling with $d$ is one of the main motivations to opt for gradient-based MCMC algorithms, such as the MALA.

We constantly monitor the acceptance rate and adjust the step size $s$ to achieve the algorithm's optimal acceptance rate.
To this end, the acceptance rate of the previous interval is determined every 500 MCMC steps, and either increased or decreased by $10\%$, if the acceptance rate is below $0.2$ or above $0.3$, respectively.
If the rate is within the desired range, the step size is kept constant.

\subsubsection{Metropolis-Adjusted Langevin Algorithm}\label{subsec:metropolis-adjusted-langevin-algorithm}

\noindent
The Langevin algorithm~\cite{Roberts2002} derives its proposal mechanism from a discretised Langevin diffusion.
It can be seen as a random walk with a drift term that is proportional to the gradient of the log-density.
While the continuous Langevin diffusion has the targeted log-density as its stationary distribution, its discretisation introduces a bias.
The Metropolis-adjusted Langevin algorithm (MALA) corrects for this bias by adding a Metropolis-Hastings correction to the mechanism.

The proposal mechanism of the MALA is given by

\begin{align}
    \bs{\theta}^* = \bs{\theta}^n + \frac{s^2}{2} \bs{M} \nabla \ln f(\bs{\theta}^n) + s \sqrt{\bs{M}}\bs{z}^n.
    \label{eq:mala_proposal}
\end{align}

\noindent
which gives rise to the Gaussian proposal distribution

\begin{align}
    q(\bs{\theta}^*|\bs{\theta}^n) = \mathcal{N} \left(\bs{\theta}^*| \bs{\theta}^n + \frac{s^2}{2} \bs{M} \nabla \ln f(\bs{\theta}^n), s^2 \bs{M} \right).
\end{align}

\noindent
The MALA proposal is similar to the RWM proposal \cref{eq:gaussian_proposal}, but has the additional drift term which is proportional to the gradient of the unnormalised log-density $\ln f(\bs{\theta}^n)$.
Note that proposal distribution is non-symmetric in $\bs{\theta}^n$ and $\bs{\theta}^*$.
Hence, the Metropolis-Hastings ratio does not simplify to the ratio of the densities and has to be computed according to the general formula in \cref{eq:metropolis-hastings_ratio}.

As for the RWM algorithm, the preconditioner matrix $\bs{M}$ is set to the sample covariance after the first 5000 MCMC-steps.
Similarly, the step size $s$ is adaptively tuned during runtime to achieve the optimal acceptance rate of $0.574$~\cite{Roberts1998}.
Only, the target interval for the acceptance rate is set to $[0.55, 0.6]$ for the MALA.
The MALA is known to be more efficient than the RWM algorithm for high-dimensional state spaces, as it can exploit the gradient information to make larger steps in the direction of the mode of the target distribution.
 \subsection{Gaussian Process Surrogate Model}\label{subsec:gaussian-process-surrogate-model-theta}

\noindent
We employ a GP surrogate model to approximate the likelihood function $\mathcal{L}(\bs{\theta})$ in order to reduce the computational cost associated with likelihood evaluations in the many MCMC iterations.
GP regression is a non-parametric Bayesian regression method that can be used to model complex, non-linear relationships between input and output variables.
A parametric model assumes a specific functional form for the relationship between input and output variables and encodes all the information in a set of learnable parameters.
In contrast, the non-parametric GP regression model uses the data directly to make predictions.
Instead of defining distributions over the parameters of the predefined basis functions, a GP defines a distribution over functions directly.
From a mathematical angle, a GP is a collection of random variables, any finite number of which have a joint Gaussian distribution:

\begin{align}
    \bs{\mathcal{L}} \sim \mathcal{N} \left( \bs{\mathcal{L}} | \bs{0}, \bs{K}(\bs{\Theta},\bs{\Theta}) \right),
\end{align}

\noindent
where a zero mean is assumed before seeing any data.
The covariance matrix $\bs{K}(\bs{\Theta},\bs{\Theta})$ is the collection of the kernel function on all input pairs, i.e. $K_{ij} = k(\bs{\theta}_i,\bs{\theta}_j)$
The matrix specifies the similarities in the output $\bs{\mathcal{L}}$ based on their input $\bs{\Theta}$.
The kernel in this case is chosen as the squared exponential kernel~\cite{Rasmussen2006}, given by

\begin{align}
    k(\bs{\theta},\bs{\theta}') = \sigma_{\mathrm{f}}^2 \exp \left( -\frac{1}{2 \ell^2} \parallel \bs{\theta} - \bs{\theta}' \parallel^2 \right).
\end{align}

\noindent
The GP variance $\sigma_{\mathrm{f}}^2$ controls the magnitude of the fluctuations around the mean function, and its length scale $\ell$ controls the smoothness of the function.
We assume the relationship of input and output of our training data to be of the form $\eta_i = \mathcal{L}(\bs{\theta}_i) + \epsilon_i$, where $\mathcal{L}(\bs{\theta})$ is the likelihood function and $\epsilon_i$ is i.i.d. noise with variance $\sigma_{\mathrm{n}}^2$.
Although the training data originates from noise-free simulations, we admit a noise term $\sigma_{\mathrm{n}}^2 \bs{I}$ for numerical stability and model flexibility.
We denote $\bs{K}(\bs{\Theta},\bs{\Theta}^*)$ the covariance matrix between the training and test points, and $\bs{K}(\bs{\Theta}^*,\bs{\Theta}^*)$ the covariance matrix between the test points.
We can then write the joint distribution over the training output $\bs{\eta}$ and new predictions $\bs{\mathcal{L}}^*$ at locations $\bs{\Theta}^*$ as

\begin{align}
    \begin{bmatrix}
        \bs{\eta} \\
        \bs{\mathcal{L}}^*
    \end{bmatrix}
    \sim
    \mathcal{N} \left(
    \bs{0},
    \begin{bmatrix}
        \bs{K}(\bs{\Theta},\bs{\Theta}) + \sigma_{\mathrm{n}}^2 \bs{I} & \bs{K}(\bs{\Theta},\bs{\Theta}^*) \\
        \bs{K}(\bs{\Theta}^*,\bs{\Theta}) & \bs{K}(\bs{\Theta}^*,\bs{\Theta}^*)
    \end{bmatrix}
    \right).
\end{align}

\noindent
Applying Bayes' formula, the predictive distribution of the test output $\bs{\mathcal{L}}^* = \bs{\mathcal{L}}(\bs{\Theta}^*)$ given the training data $\mathcal{D}_{\mathrm{GP}} = (\bs{\Theta}, \bs{\eta} )$ is then given by

\begin{align}
    \bs{\mathcal{L}}^* | \bs{\eta}, \bs{\Theta}, \bs{\Theta}^* &\sim \mathcal{N} \left( \bs{\mathcal{L}}^* | \bs{\mu}^*, \bs{\Sigma}^* \right),\\
    \bs{\mu}^* &= \bs{K}(\bs{\Theta}^*,\bs{\Theta}) \left( \bs{K}(\bs{\Theta},\bs{\Theta}) + \sigma_{\mathrm{n}}^2 \bs{I} \right)^{-1} \bs{\eta},\nonumber \\
    \bs{\Sigma}^* &= \bs{K}(\bs{\Theta}^*,\bs{\Theta}^*) - \bs{K}(\bs{\Theta}^*,\bs{\Theta}) \left( \bs{K}(\bs{\Theta},\bs{\Theta}) + \sigma_{\mathrm{n}}^2 \bs{I} \right)^{-1} \bs{K}(\bs{\Theta},\bs{\Theta}^*). \nonumber
\end{align}

\noindent
The predictive mean $\bs{\mu}^*$ is our best guess about the function values at the test points, and the predictive variance $\bs{\Sigma}^*$ quantifies our uncertainty about these predictions.

We determine the GP model hyperparameters---the process variance $\sigma_{\mathrm{f}}^2$, the noise variance $\sigma_{\mathrm{n}}^2$, and the length scale $\ell$---via empirical Bayes for computational efficiency.
For this, we maximise the marginal likelihood of the training outputs $\bs{\eta}$ given the training input $\bs{\Theta}$.
The marginal likelihood $p(\bs{\eta} | \sigma_{\mathrm{f}}^2, \sigma_{\mathrm{n}}^2, \ell)$ and its logarithm $L$ are given by

\begin{align}
    p(\bs{\eta} | \sigma_{\mathrm{f}}^2, \sigma_{\mathrm{n}}^2, \ell)
    &= \int p(\bs{\eta} | \bs{\mathcal{L}}, \sigma_{\mathrm{n}}^2) p(\bs{\mathcal{L}} | \sigma_{\mathrm{f}}^2, \ell) d\bs{\mathcal{L}},
    \label{eq:marginal_likelihood}\\
    L &= \ln p(\bs{\eta} | \sigma_{\mathrm{f}}^2, \sigma_{\mathrm{n}}^2, \ell).\label{eq:log_marginal_likelihood}
\end{align}

\noindent
Instead of maximising the marginal likelihood directly, we maximise the log marginal likelihood \cref{eq:log_marginal_likelihood} to avoid numerical issues.
The associated optimisation problem is solved with the BFGS algorithm~\cite{Fletcher2000}.
Due to the non-convex nature of the optimisation problem, the optimisation is performed multiple times with different initial guesses for the hyperparameters to ensure convergence to the global or a sufficiently good local maximum.
The obtained point estimates for the hyperparameters are then used for the subsequent prediction. 
\section{MCMC guided active learning}\label{sec:mcmc_guided_active_learning}

\noindent
In this section, we introduce an algorithm that employs MCMC-guided sampling as a straightforward active learning strategy for constructing surrogate models.
This approach leverages the predictive uncertainty inherent in the GP surrogate model, coupled with the trajectory of the MCMC algorithm, drawing inspiration from the active learning framework presented in~\citet{Rocha2021}.

Initially, we generate a set of $N_0$ samples from the posterior distribution using the forward model.
This set forms the initial training dataset for our GP surrogate model.
Upon this dataset, we fit the GP model by estimating its hyperparameters and record the initial log marginal likelihood, denoted as $L_{\mathrm{old}}$.
After the initialisation phase is completed, the MCMC algorithm generates new proposals based on the GP model.
Denoting $\bs{k}(\bs{\theta}^*)$ the covariance vector between the training data and the new point $\bs{\theta}^*$, and $\bs{K}$ the covariance matrix of the training data, the predictive variance of the GP model at $\bs{\theta}^*$ is given by

\begin{align}
    \mathbb{V}[\mathcal{L}(\bs{\theta}^*)] &= k(\bs{\theta}^*, \bs{\theta}^*) - \bs{k}(\bs{\theta}^*)^\top\left(\bs{K} + \sigma^2\bs{I}\right)^{-1}\bs{k}(\bs{\theta}^*) \label{eq:gp_predictive_point_variance_1}.
\end{align}

\noindent
If this variance remains below a specified threshold $\gamma_{\mathrm{v}}$, we accept the GP's prediction at $\bs{\theta}^*$ and proceed without further action.
Conversely, if the predictive variance exceeds the variance threshold $\gamma_{\mathrm{v}}$, we resort to evaluating the forward model $\bs{y}^* = \mathcal{M}(\bs{\theta}^*)$ and compute the likelihood $\mathcal{L}(\bs{\theta}^*)$ to obtain a new observation.
We will refer to $\gamma_{\mathrm{v}}$ as the reject threshold.
This observation is then used to update our training dataset.
Following the addition of new data, we update the GP's posterior covariance matrix and recalculate the log marginal likelihood, denoted as $L_{\mathrm{new}}$.
If the ratio $\left| \nicefrac{L_{\mathrm{new}}}{L_{\mathrm{old}}} \right|$ exceeds a predetermined threshold $\gamma_{\mathrm{L}}$, this indicates a significant change in the model's understanding, prompting a re-estimation of the GP hyperparameters based on the updated dataset.
We will thus refer to $\gamma_{\mathrm{L}}$ as retrain threshold.
The process iterates until the end of the burn-in phase $N_{\mathrm{b}}$ is reached, at which point the GP model's hyperparameters are fixed.
These parameters have a significant impact on the model's predictive accuracy and their choice is therefore investigated in \cref{sec:active-learning-validation}.

Replacing the forward model with the GP surrogate model leads to a modified proposal distribution \cref{eq:mala_proposal} for the MALA:

\begin{align}
    \tilde{p}(\bs{\theta}^* | \bs{\theta}) = \mathcal{N} \left( \bs{\theta}^* | \bs{\theta}^n + \frac{s^2}{2} \bs{M}^{-1} \nabla_{\bs{\theta}}
    \left( \mathbb{E} \left[ \mathcal{L}(\bs{\theta}^n) \right]+ \ln p(\bs{\theta}^n) \right), s^2 \bs{M}\right).
    \label{eq:modified-mala-proposal}
\end{align}

\noindent
The modified MALA acceptance probability is then given by:

\begin{align}
    \tilde{a}(\bs{\theta}^n, \bs{\theta}^*) = \min \left( 1,
    \frac{\exp \left (\mathbb{E}\left[\mathcal{L}(\bs{\theta}^*)\right] \right)   p(\bs{\theta}^*)   \tilde{q}(\bs{\theta}^*|\bs{\theta}^{n})}
    {\exp \left( \mathbb{E} \left[\mathcal{L}(\bs{\theta}^{n})\right] \right)  p(\bs{\theta}^n)   \tilde{q}(\bs{\theta}^{n}|\bs{\theta}^*)}
    \right)
    \label{eq:modified-mala-acceptance}
\end{align}

\noindent
The proposal for the RWM algorithm \cref{eq:gaussian_proposal} remains the same, as it does not depend on the forward model.
However, the acceptance probability \cref{eq:acceptance_probability} is modified to account for the GP surrogate model:

\begin{align}
    \tilde{a}(\bs{\theta}^n, \bs{\theta}^*) = \min \left( 1,
    \frac{\exp \left (\mathbb{E}\left[\mathcal{L}(\bs{\theta}^*)\right] \right)   p(\bs{\theta}^*)}
    {\exp \left( \mathbb{E} \left[\mathcal{L}(\bs{\theta}^{n})\right] \right)  p(\bs{\theta}^n)}
    \right)
    \label{eq:modified-rwm-acceptance}
\end{align}

\noindent
The MCMC algorithm incurs a bias from the surrogate model in the acceptance probabilities \cref{eq:modified-mala-acceptance,eq:modified-rwm-acceptance}.
The quantification of this bias and its impact on the posterior estimation is at the heart of this work.

As the GP model refines its accuracy with the incorporation of new data, we anticipate a reduction in predictive variance, signaling an enhanced model fidelity.
Consequently, the need to resort to the computationally expensive forward model diminishes.
This strategy ensures that new data points are strategically added to the model's training set, particularly focusing on regions yet to be explored by the MCMC algorithm, without changing the response surface in regions that have been visited in previous steps.
The active learning strategy is described in detail in~\cref{alg:active_learning}.

\begin{algorithm}
    \begin{algorithmic}[1]
    \State Initialise $N_0$ samples from the posterior using the forward model to obtain initial dataset $\mathcal{D}_{\mathrm{GP}} = \{(\bs{\theta}_0, \mathcal{L}(\bs{\theta}_0)), \dots, (\bs{\theta}_N, \mathcal{L}(\bs{\theta}_{N_0}))\}$
    \State Estimate hyperparameters on $\mathcal{D}_{\mathrm{GP}}$ and record $L_{\mathrm{old}}$
    \State $n \gets 0$
    \While{$n < N_{\mathrm{total}}$}
        \State Generate new point $\bs{\theta}^*$ from the proposal distribution $\tilde{q}(\bs{\theta}^*|\bs{\theta}^n)$ \label{alg:proposal}
        \If{$\mathbb{V}[\mathcal{L}(\bs{\theta}^*)] < \gamma_{\mathrm{v}}$}
            \State Accept GP prediction for $\mathcal{L}(\bs{\theta}^*)$
        \Else
            \State Evaluate forward model $\bs{y}^* = \mathcal{M}(\bs{\theta}^*)$, compute likelihood $\mathcal{L}(\bs{\theta}^*)$
            \State Update training dataset $\mathcal{D}_{\mathrm{GP}} \gets \mathcal{D}_{\mathrm{GP}} \cup \left( \bs{\theta}^*, \mathcal{L}(\bs{\theta}^*) \right)$
            \State Update GP's posterior covariance matrix and compute $L_{\mathrm{new}}$
            \If{$\left| \nicefrac{L_{\mathrm{new}}}{L_{\mathrm{old}}} \right| > \gamma_{\mathrm{L}}$ AND $n < N_{\mathrm{b}}$}
                \State Re-estimate GP hyperparameters and update $L_{\mathrm{new}}$
                \State $L_{\mathrm{old}} \gets L_{\mathrm{new}}$
            \EndIf
        \EndIf
        \If{$\tilde{a}(\bs{\theta}^n, \bs{\theta}^*) > u \sim \mathcal{U}(0,1)$} \label{alg:acceptance}
            \State $\bs{\theta}^{n+1} \gets \bs{\theta}^*$
        \Else
            \State $\bs{\theta}^{n+1} \gets \bs{\theta}^n$
        \EndIf
        \State $n \gets n + 1$
    \EndWhile
    \end{algorithmic}
    \caption{MCMC-guided active learning for GP surrogate model construction.
    Depending on the choice of the algorithm, the proposal density $\tilde{q}(\bs{\theta}^*|\bs{\theta}^n)$ in \cref{alg:proposal} is either given by \cref{eq:gaussian_proposal} for the RWM algorithm or \cref{eq:modified-mala-proposal} for the MALA.
    Similarily, the acceptance probabiliy $\tilde{\alpha}$ in \cref{alg:acceptance} is given by \cref{eq:modified-rwm-acceptance} (RWM) or \cref{eq:modified-mala-acceptance} (MALA).}
    \label{alg:active_learning}
\end{algorithm} \section{Case study}\label{sec:case_study}

\begin{figure}
    \centering
    \def\svgwidth{0.7\textwidth}
    \begingroup \makeatletter \providecommand\color[2][]{\errmessage{(Inkscape) Color is used for the text in Inkscape, but the package 'color.sty' is not loaded}\renewcommand\color[2][]{}}\providecommand\transparent[1]{\errmessage{(Inkscape) Transparency is used (non-zero) for the text in Inkscape, but the package 'transparent.sty' is not loaded}\renewcommand\transparent[1]{}}\providecommand\rotatebox[2]{#2}\newcommand*\fsize{\dimexpr\f@size pt\relax}\newcommand*\lineheight[1]{\fontsize{\fsize}{#1\fsize}\selectfont}\ifx\svgwidth\undefined \setlength{\unitlength}{595.27559055bp}\ifx\svgscale\undefined \relax \else \setlength{\unitlength}{\unitlength * \real{\svgscale}}\fi \else \setlength{\unitlength}{\svgwidth}\fi \global\let\svgwidth\undefined \global\let\svgscale\undefined \makeatother \begin{picture}(1,0.20952381)\lineheight{1}\setlength\tabcolsep{0pt}\put(0,0){\includegraphics[width=\unitlength,page=1]{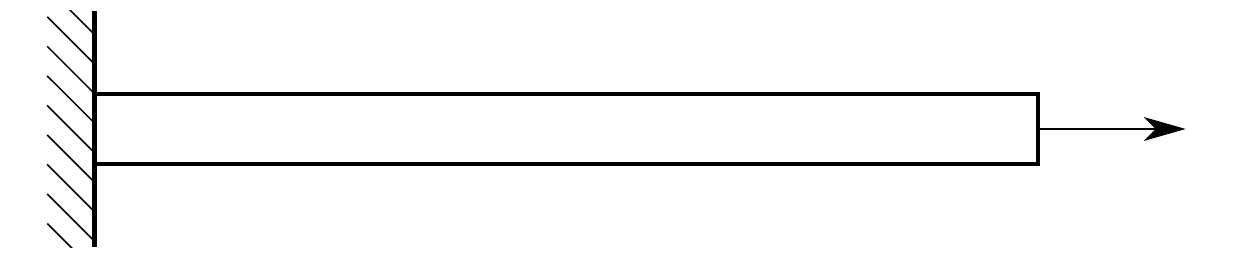}}
\put(0.7955562,0.15843264){\makebox(0,0)[lt]{\lineheight{1.25}\smash{\begin{tabular}[t]{l}$u(1,t)=\hat{u}(1,t)$\end{tabular}}}}\put(0.10109121,0.15843264){\makebox(0,0)[lt]{\lineheight{1.25}\smash{\begin{tabular}[t]{l}$u(0,t)=0$\end{tabular}}}}\end{picture}\endgroup      \caption{Schematic representation of the one-dimensional bar problem. Dirichlet boundary conditions are enforced at the left and right boundaries.}
    \label{fig:1d_bar_schematic}
\end{figure}

\noindent
In this section, we introduce a scalable case study to elucidate the effects of the methodological decision on surrogate modelling and sampling.
First and foremost, the problem must be computationally inexpensive.
Since we cannot obtain an analytical posterior, we need the capability to brute-force approximate this distribution via long runs of MCMC to obtain a reference solution.
Additionally, the problem should closely resemble a real-world scenario.
To this end, we introduce a one-dimensional bar with length $L = 1$ and constant cross-sectional area $A = 1$, depicted in figure \cref{fig:1d_bar_schematic}.

\subsection{Material Model}

\noindent
To make things more challenging, we introduce a simple non-linear constitutive model and a spatially varying initial stiffness.
It is adopted as a fast and pragmatic choice, but nevertheless without loss of generality.
The constitutive model is given by

\begin{align}
    \sigma (\varepsilon) = r (E - H) \left( 1 - \mathrm{exp} \left(-\frac{\varepsilon}{r} \right) \right) + H \varepsilon.   \label{eq:expl_material_model}
\end{align}

\noindent
This model, characterised by its initial stiffness $E$, its terminal stiffness $H$, and its rate parameter $r$, mimics nonlinear material behaviour during plastic deformation.
It does so without necessitating the computational burden associated with tracking the evolution of plastic strain.
Note that this simplification is only reasonable for the monotonic loading considered in this study.
We thereby embed additional non-linearity into the system without the exhaustive computational requirements typical of a complete plasticity model.

\begin{figure}
    \centering
    \includegraphics[width=.8\textwidth]{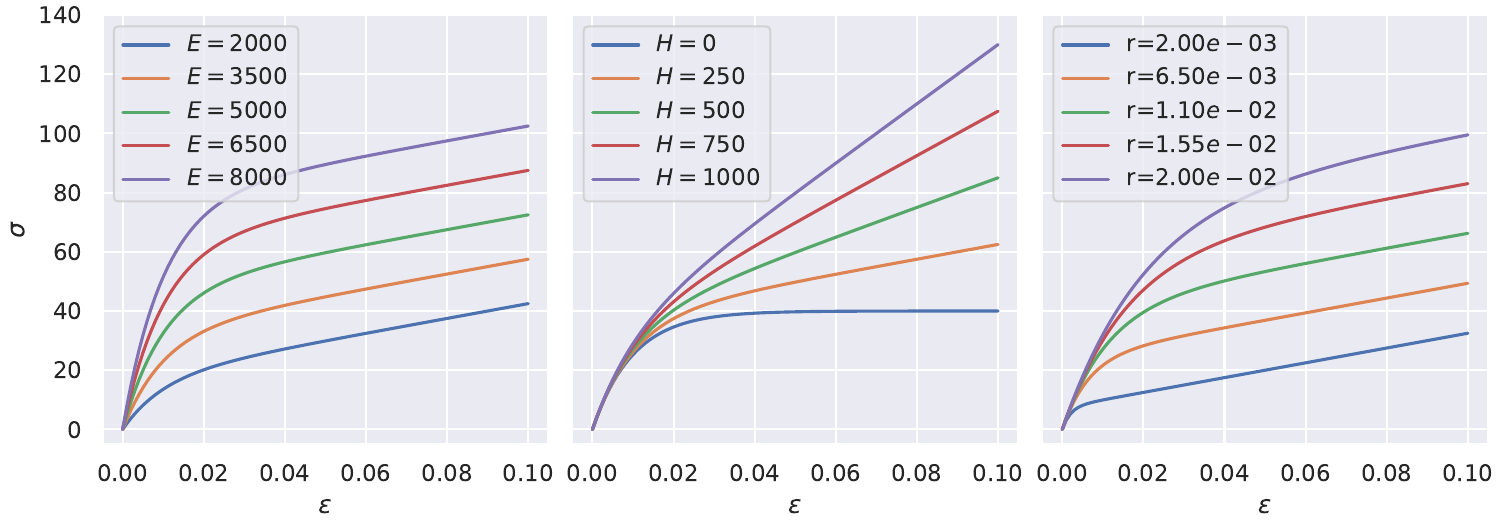}
    \caption{Non-linear material model with parameters $E = 4000$, $H = 250$, and $r = 0.01$, unless specified otherwise. The three subplots show the stress-strain relationship for different values of the material parameters.}
    \label{fig:material_model}
\end{figure}

\cref{fig:material_model} illustrates the stress-strain relationship for different values of the material parameters.
The material model's non-linearity is evident, with the stress-strain curve exhibiting a distinct hardening behaviour.
After initial numerical testing, no single parameter stood out as comparatively more challenging to infer, except for a hardening modulus $H$ close to zero.
We, therefore, focus exclusively on the initial stiffness $E$ in the subsequent analysis to ensure a consistent comparison across the different methods at fixed $H$ and $r$.
The spatially varying initial stiffness $\hat{\beta} = \hat{E} = \hat{E}(x) \approx E(x, \bs{\theta})$ is hence the unknown field to be inferred.

\subsection{Sequence of Inverse Problems}

\begin{figure}
    \centering
    \subcaptionbox{$d = 2$ \label{fig:problem_sequence_d2}}[0.49\textwidth]{
        \includegraphics[width=\linewidth]{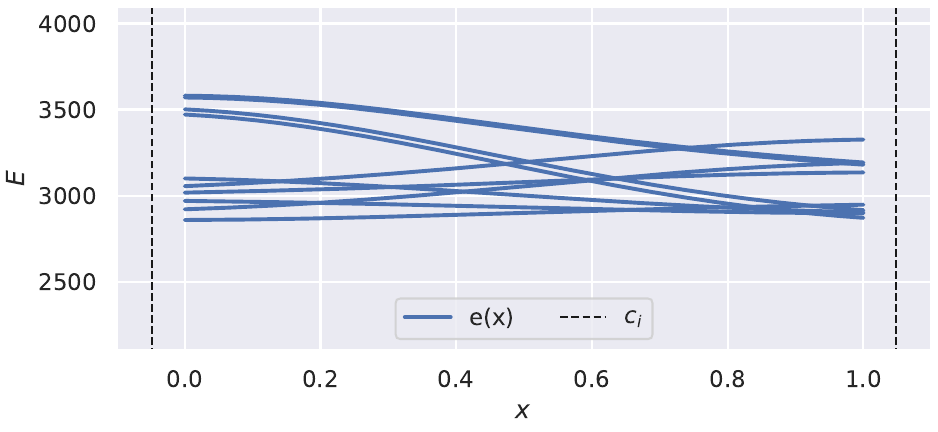}
    }
    \subcaptionbox{$d = 5$\label{fig:problem_sequence_d5}}[0.49\textwidth]{
        \includegraphics[width=\linewidth]{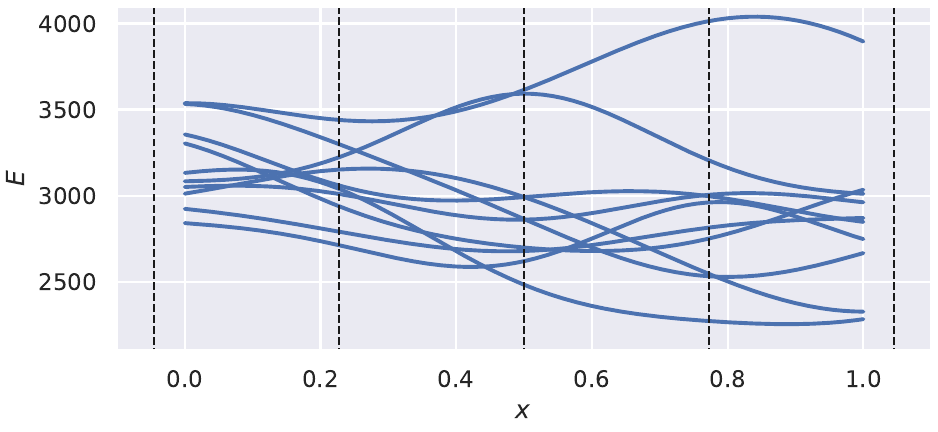}
    }
    \\ \vspace{0.1cm}
    \subcaptionbox{$d = 10$\label{fig:problem_sequence_d10}}[0.49\textwidth]{
        \includegraphics[width=\linewidth]{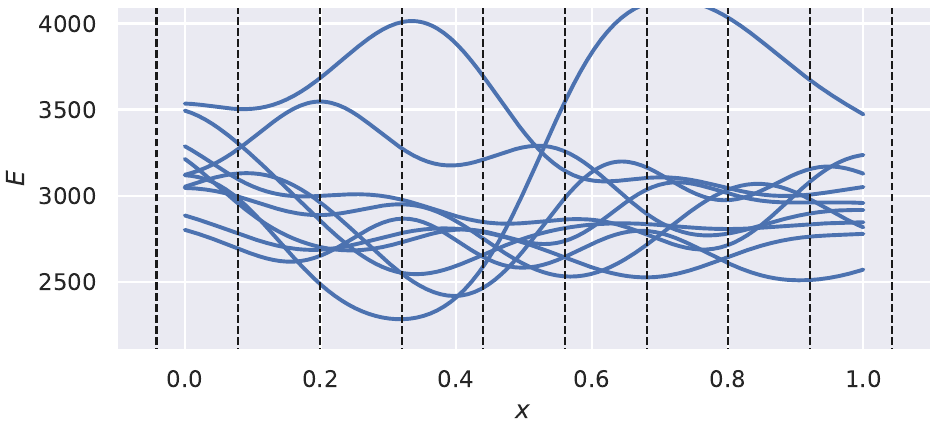}
    }
    \subcaptionbox{$d = 20$\label{fig:problem_sequence_d20}}[0.49\textwidth]{
        \includegraphics[width=\linewidth]{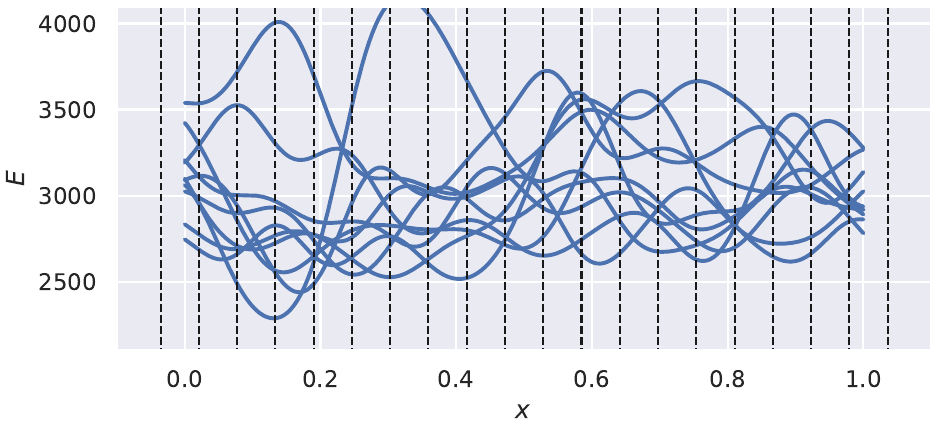}
    }
    \caption{Random field realisations of the spatially varying initial stiffness for different numbers of RBFs and respective length scales. The blue lines depict different generations of the ground truth, which is sampled from the prior distribution. The dashed black lines represent the centers of the RBFs.}
    \label{fig:problem_sequence}
\end{figure}

\noindent
We define a sequence of inverse problems by considering different stochastic discretisations of the spatially varying initial stiffness.
Taking the spatial correlation of the RBF discretisation into account, further additions of basis functions are expected to yield diminishing gains of information for a given random field.
As the RBFs with centers close to one another are expected to correlate, their inference will be less challenging.
We adapt the length scale of the random field, and thereby the complexity of the inference problem, according to the number of terms in the random field expansion, as demonstrated in \cref{fig:problem_sequence}.
In other words, we must infer a random function with an expected number of zero crossings proportional to the number of free parameters.
We also increase the amount of displacement data to ensure that the observations-to-parameters ratio remains constant.
This ensures any difference in inference performance as the dimensionality of the latent space increases can be ascribed to the dimensionality and not to the complexity of the inferred behaviour.

\begin{figure}
    \centering
    \includegraphics[width=0.6\textwidth]{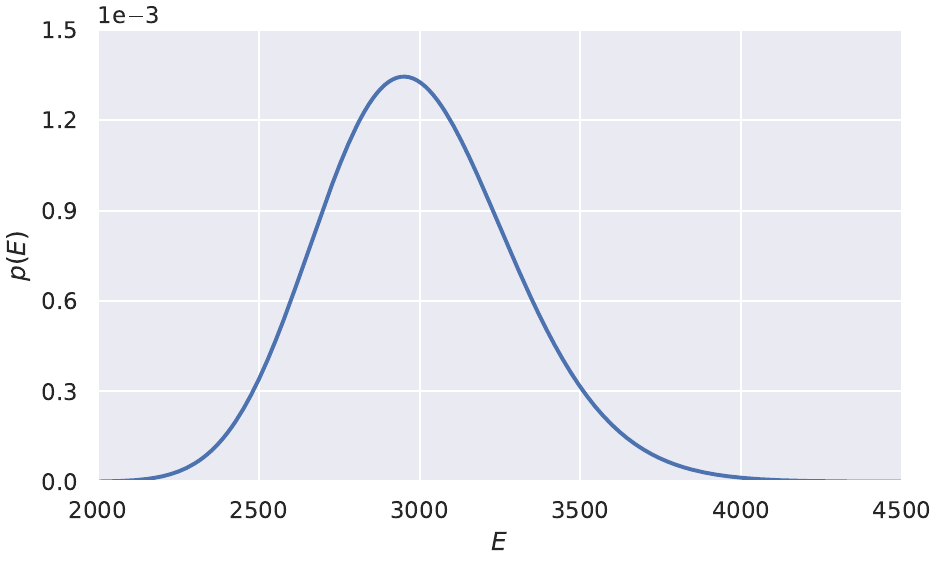}
    \caption{Marginal of the log-normal field $E(x)$.}
    \label{fig:log_normal_marginal}
\end{figure}

As we know that the initial stiffness must be positive, we apply an exponential transformation to a random field $\hat{\bar{E}}(x)$.
If this random field $\hat{\bar{E}}(x)$ is Gaussian, the exponential transformation results in a log-normal distribution.
The relationships between the log-normal field $\hat{E}(x)$, its discretised version $E(x, \bs{\theta})$, and the respective underlying Gaussian fields $\hat{\bar{E}}(x)$ and $\bar{E}(x, \bs{\theta})$ are given by

\begin{align}
    \hat{E}(x) &= \exp \left( \hat{\bar{E}}(x) \right),\label{eq:log-normal-field}\\
    E(x, \bs{\theta}) &= \exp (\bar{E}(x, \bs{\theta})) = \exp \left( \mu_{\bar{E}}(x) + \sum_{i=1}^{d} \phi_i (x) \theta_i \right) \label{eq:log-normal-field-discretized},
\end{align}

\noindent
where $\mu_{\tilde{E}}(x)$ is the mean of the underlying normal distribution.
We set the random field parameters to $\mu_{\bar{E}}(x) = 8$, $\sigma_{\theta} = 0.1$ to mimic an elastoplastic polymer used as matrix material in fiber reinforced composites~\cite{Melro2013}, and fix $l = \nicefrac{1.5}{d}$.
The corresponding marginal distribution is depicted in \cref{fig:log_normal_marginal} and the random field realisations for different numbers of RBFs and respective length scales are shown in \cref{fig:problem_sequence}.
The goal is then to find the posterior distribution of the coefficients $\bs{\theta}$ of the underlying Gaussian random field, which can then be transformed to the posterior of the parameter field with help of \cref{eq:log-normal-field-discretized}.
We fix the remaining parameters to $H = 100$ and $r = 0.01$.

In our fictitious experiment, we enforce dirichlet boundary conditions on the left and right boundaries as follows:

\begin{align}
    u(0,t) &= 0, \label{eq:dirichlet_bc_left} \\
    u(1,t) &= t \quad \mathrm{for} \quad t \in [0,T]. \label{eq:dirichlet_bc_right}
\end{align}

\noindent
The displacement of the right boundary increases linearly with pseudo time $t$ and the final time is set to $T = 0.1$.
We sample the ground truth from the prior distribution (see \cref{eq:rbf_variance,eq:log-normal-field-discretized} and \cref{fig:problem_sequence}) and solve the forward problem.
To generate the observations, we record the displacements at a number of locations and the reaction force at the far right boundary at $N_j = 5$ distinct times $t_{j} = \{0.02, 0.04, 0.06, 0.08, 0.1\}$.
The number of recorded displacements per time $t_j$ is set to $N_i = [\nicefrac{3d}{4}]$, where $[\cdot]$ denotes the closest integer.
This means there are fewer measurement locations $x_i$ than parameters to infer.
These locations are spread uniformly within the domain $\mathscr{D}$.

We then synthesise measurements from the recorded displacements and forces according to our observation model \cref{eq:likelihood}.
We assume uncorrelated Gaussian noise, i.e. $\bs{\Sigma}_{u\!u} = \sigma_u^2 \bs{I}$ and $\bs{\Sigma}_{f\!f} = \sigma_f^2 \bs{I}$, with respective standard deviations $\sigma_{u} = 0.001$ and $\sigma_{f} = 1$.
This virtual experiment is conducted only once per given ground truth.
The displacements and forces are stacked, i.e. $\bs{u}=[u(x_0, t_0), u(x_1, t_0), \cdots, u(x_{N_i}, t_{N_j})]$ and $\bs{f} = [f_{0}, \dots, f_{N_j}]$.
The full observation vector $\tilde{\bs{y}} = [\bs{u}^\mathsf{T}, \bs{f}^\mathsf{T}]^\mathsf{T}$ then comprises a total of $m = N_j \cdot (N_i + 1)$ scalar observations.
The goal is now to infer the parameters of a matching discretisation of the random field from these noisy observations.

\subsection{Wasserstein Distance}

\noindent
So far, we have introduced a scalable case study, two MCMC algorithms (RWM and MALA) to solve the inverse problems, and a GP surrogate model that can facilitate this process.
To understand the impact of the methodological decisions on the inference process, we must compare the resulting approximate posterior distributions.
However, neither our ground truth nor the surrogate-model-assisted inference are available in analytical form, but instead are represented with samples from MCMC simulations.
We must therefore find a suitable metric to compare these empirical distributions.
The Wasserstein distance between two distributions is the minimum effort required to morph one into the other.
This metric is particularly advantageous for empirical distributions, such as those represented by MCMC samples, because it can be directly applied to point clouds without requiring the distributions to be expressed in a functional form.

Other metrics, such as the Kullback-Leibler divergence or the total variation distance inherently require at least one of the distributions to be functionally described.
To use these metrics in the given settings, one must approximate the empirical distribution using kernel density estimation or another density estimation technique, which introduces additional complexity and uncertainty into the analysis.
In contrast, the Wasserstein distance directly considers the actual distances between points avoiding the pitfalls associated with density estimation.
For our case study, we specifically employ the Wasserstein 2-distance, which is the square root of the optimal transport cost, and denote it as $W_2$ in the following.

The Wasserstein 2-distance $W_2(P, Q)$ between two continuous distributions $P$ and $Q$ on a metric space $\mathcal{X}$ is defined as:

\begin{align}
W_2(P, Q) = \left( \inf_{\gamma \in \Gamma(P, Q)} \int_{\mathcal{X} \times \mathcal{X}} \|x - y\|^2 \, d\gamma(x, y) \right)^{\frac{1}{2}},
\end{align}

\noindent
where $\Gamma(P, Q)$ is the set of all joint distributions on $\mathcal{X} \times \mathcal{X}$ with marginals $P$ and $Q$.
In case of empirical distributions represented by an equal amount of samples $N_{\mathrm{total}}$, the Wasserstein 2-distance reduces to:

\begin{align}
W_2(P, Q) = \left( \frac{1}{N_{\mathrm{total}}} \sum_{i=1}^{N_{\mathrm{total}}} \|x_i - y_{\pi(i)}\|^2 \right)^{\frac{1}{2}},
\end{align}

\noindent
where $x_i$ and $y_i$ are the samples from the distributions $P$ and $Q$, respectively, and $\pi$ is the permutation that minimises the square distance.

 \section{Results}\label{sec:results}

\noindent
In this section, we present the results for the case study.
We first demonstrate the inference on the bar problem in general.
Next, we validate the active learning approach on the five-dimensional problem and compare it to the a priori training strategies.
Finally, we look at the impact of the MCMC algorithm on the surrogate model construction and on the efficiency and accuracy of the inference.

\subsection{Reference Solution}\label{sec:reference_solution}

\begin{figure}
    \centering
    \subcaptionbox{$d = 2$ \label{fig:posterior_d2}}[0.49\textwidth]{
        \includegraphics[width=\linewidth]{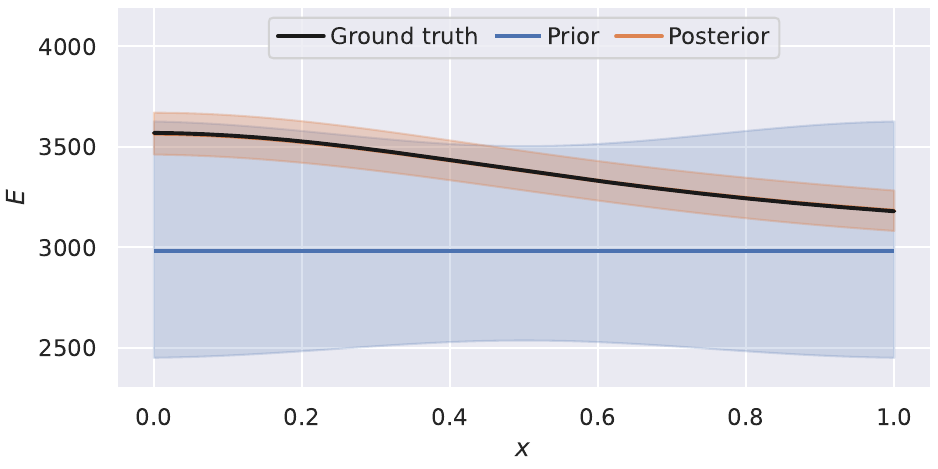}
    }
    \subcaptionbox{$d = 5$ \label{fig:posterior_d5}}[0.49\textwidth]{
        \includegraphics[width=\linewidth]{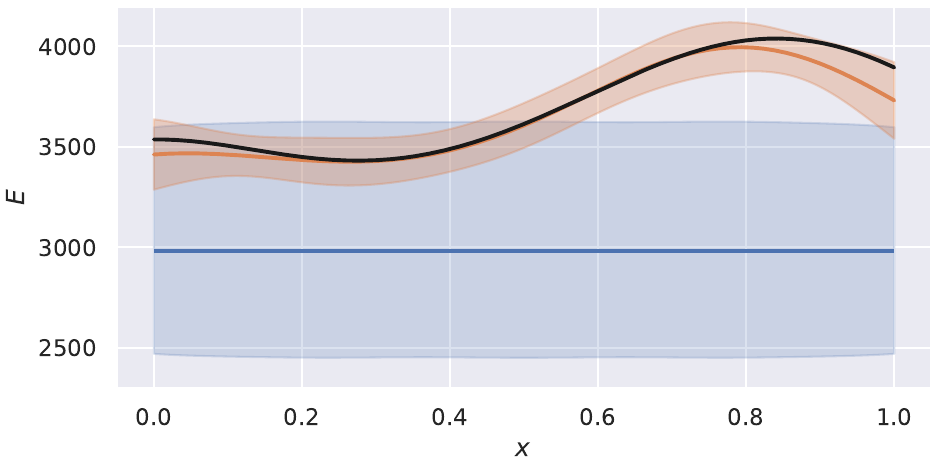}
    }
    \smallskip\\
    \subcaptionbox{$d = 10$ \label{fig:posterior_d10}}[0.49\textwidth]{
        \includegraphics[width=\linewidth]{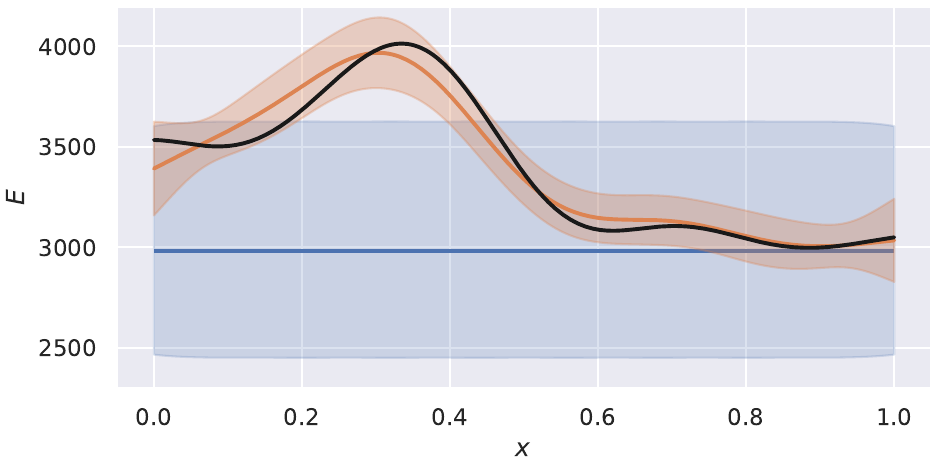}
    }
    \subcaptionbox{$d = 20$ \label{fig:posterior_d20}}[0.49\textwidth]{
        \includegraphics[width=\linewidth]{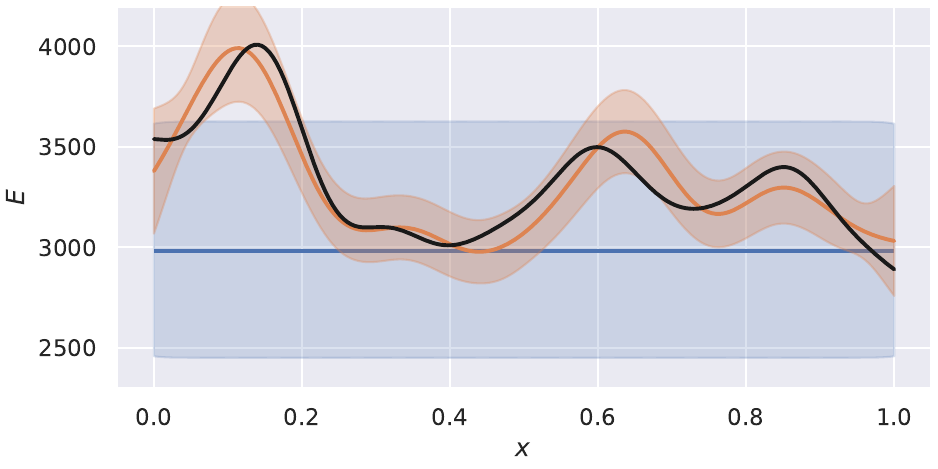}
    }
    \caption{Posterior realisations of the spatially varying initial stiffness $E$ for different numbers of RBFs and respective length scales.
    The shaded areas represent the 95\% credible intervals of their respective distribution.}
    \label{fig:posterior_fields}
\end{figure}

\noindent
We first demonstrate the inference process on the one-dimensional bar problem with varying dimensionality of the unknown field.
\cref{fig:posterior_fields} shows the ground truth---a sample from the prior---alongside the prior and posterior distributions of the initial stiffness.
All results are obtained using the RWM algorithm exclusively relying on the FEM model throughout the MCMC run.
In these and all other runs, a total of \num{200000} MCMC samples were generated for each run, with the first \num{5000} samples discarded as burn-in.
Further, each chain was thinned by a factor of 40 to ensure uncorrelated posterior samples.
This yields a total of \num{4500} samples per posterior distribution, keeping the cost of computing the Wasserstein distance manageable.
The posterior distributions of the initial stiffness are in good agreement with the ground truth in all scenarios presented in \cref{fig:posterior_fields}, underscoring the efficacy of the Bayesian approach.
Further, the true fields are well contained within the 95\% credible interval of the posterior.
This consistency indicates a robust and meaningful quantification of the uncertainty, affirming the reliability of our inference process.
With this non-linear, yet scalable sequence of inverse problems at hand, we now turn our attention to the many modelling choices in surrogate model assisted inference.

 \subsection{Active Learning Validation}\label{sec:active-learning-validation}

\noindent
Before diving into the comparison of the various training strategies, we must first validate the active learning approach.

\subsubsection{Influence of Active Learning Parameters}

\begin{figure}
    \centering
    \subcaptionbox{$\gamma_{\mathrm{v}} = 1.0$ \label{fig:reject_1.0}}[0.48\textwidth]{
        \includegraphics[width=\linewidth]{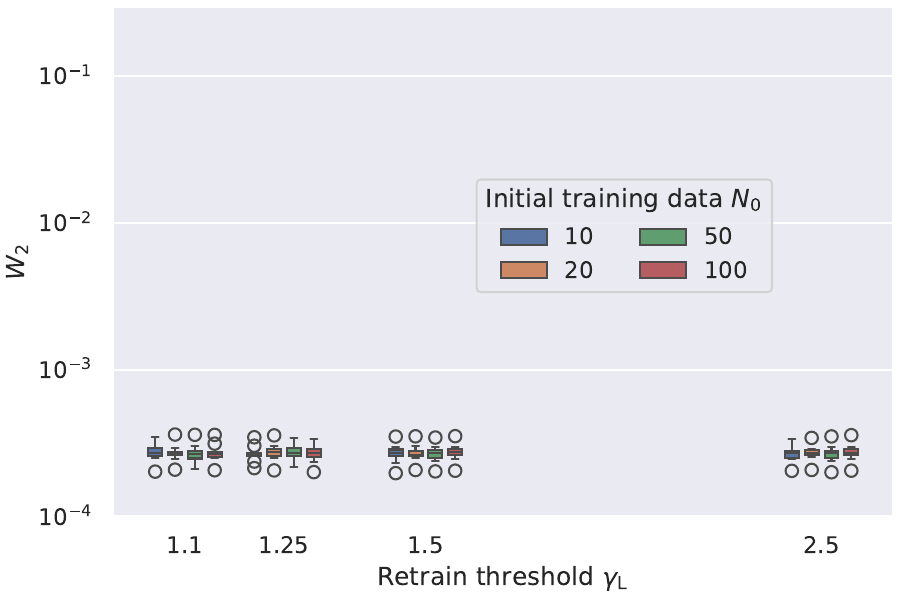}
    }
    \subcaptionbox{$\gamma_{\mathrm{v}} = 5.0$ \label{fig:reject_5.0}}[0.48\textwidth]{
        \includegraphics[width=\linewidth]{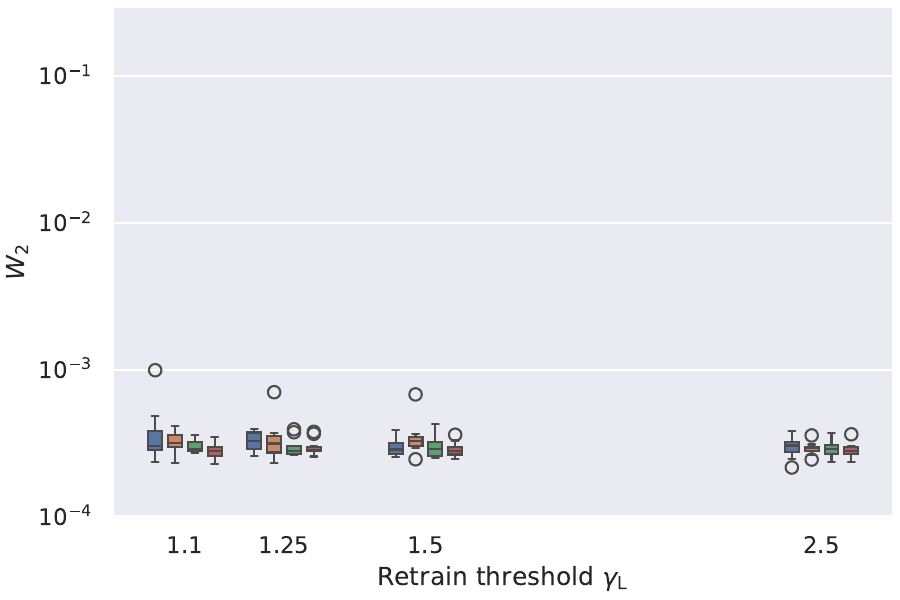}
    }
    \\ \vspace{0.1cm}
    \subcaptionbox{$\gamma_{\mathrm{v}} = 20.0$ \label{fig:reject_20.0}}[0.48\textwidth]{
        \includegraphics[width=\linewidth]{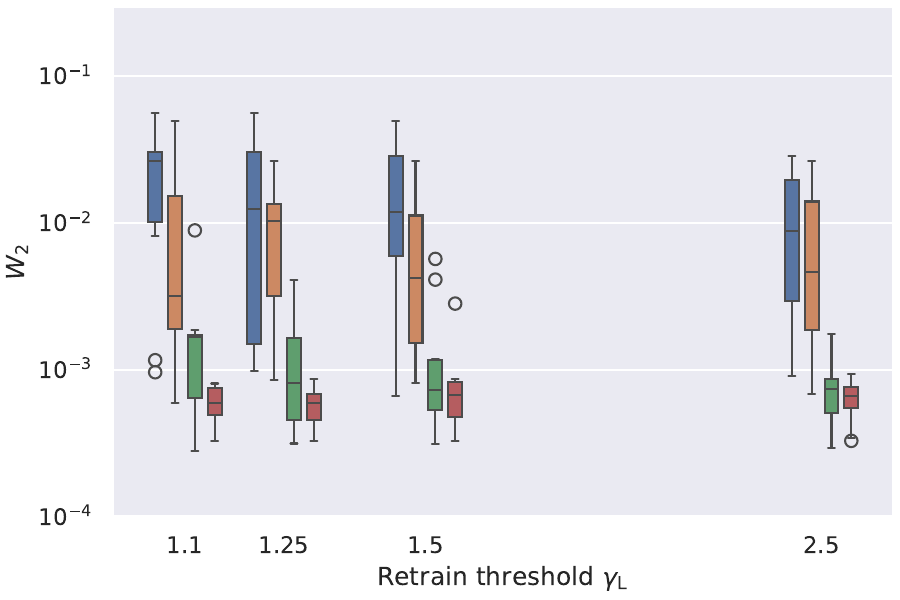}
    }
    \subcaptionbox{$\gamma_{\mathrm{v}} = 100.0$ \label{fig:reject_100.0}}[0.48\textwidth]{
        \includegraphics[width=\linewidth]{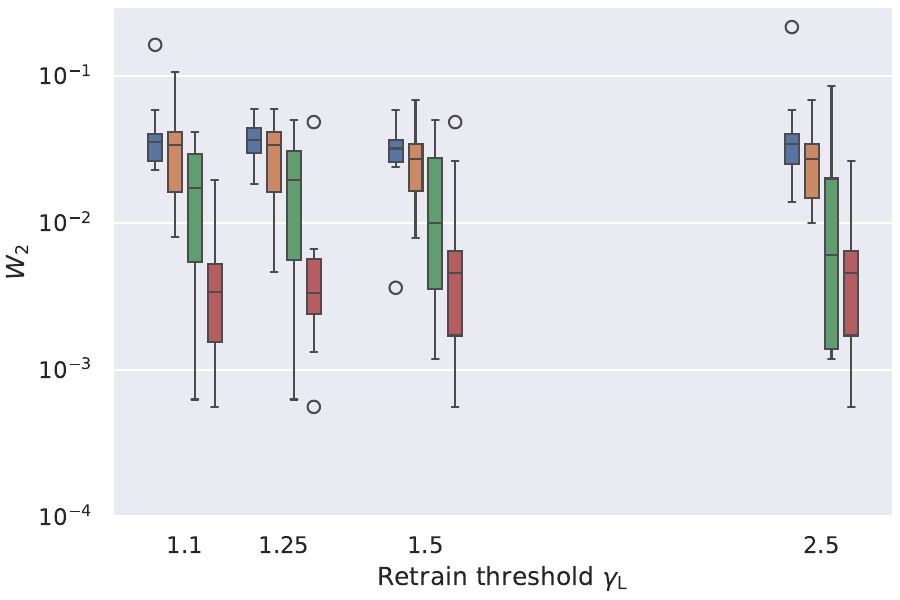}
    }
    \caption{Accuracy of the active learning approach on the 5 dimensional problem for different values of the reject threshold $\gamma_\mathrm{v}$ in subfigures (a) - (d).
    Each subfigure shows the Wasserstein 2-distance $W_2$ between the reference and approximate posterior distribution for different values of the retrain threshold $\gamma_\mathrm{L}$. The number of initial training data $N_0$ is color coded.}
    \label{fig:rejects}
\end{figure}

We use the 5-dimensional problem to study the influence of the active learning parameters.
\cref{alg:active_learning} features three parameters that require tuning: the number of pretrain steps $N_0$, the reject threshold $\gamma_\mathrm{v}$, and the retrain threshold $\gamma_\mathrm{L}$.
Every setting is repeated 10 times with different random seeds to account for the stochastic nature of the MCMC algorithm.
The GP hyperparameter optimisation is performed 200 times with different initialisations to ensure convergence to the global minimum.
The result of a grid search over these parameters is shown in \cref{fig:rejects}.
The Wasserstein 2-distance with respect to the reference solution from \cref{sec:reference_solution} is used as a metric to evaluate the accuracy of the active learning approach.
While there is little variance in the results in \cref{fig:reject_1.0} and \cref{fig:reject_5.0}, the results in \cref{fig:reject_20.0} and \cref{fig:reject_100.0} show the impact of the initial training points $N_0$:
the more training points, the smaller the approximation error.
It also becomes clear that the retrain threshold $\gamma_\mathrm{L}$ only has a minor impact on the accuracy of the inference.

The reject threshold $\gamma_\mathrm{v}$ emerges as the dominant factor when comparing the results of \cref{fig:reject_1.0} - \cref{fig:reject_100.0}, each of which represent a different value of $\gamma_\mathrm{v}$.
No matter the number of initial MCMC steps $N_0$ or the retrain threshold $\gamma_{\mathrm{L}}$, the accuracy of the active learning approach clearly depends on the reject threshold $\gamma_{\mathrm{v}}$.
These results also suggest that the active learning approach is robust with respect to the initial hyperparameter estimation, as the Wasserstein 2-distance appears to be stable across the different values of the retrain threshold.
We set the active learning parameters to $N_0 = 20$, $\gamma_\mathrm{L} = 2.5$, and $\gamma_\mathrm{v} \in \{1.0, 5.0, 20.0\}$ for the remaining experiments.

 \subsection{Online vs Offline Learning}\label{sec:online_vs_offline}

\noindent
This section explores the training strategies for surrogate models, specifically contrasting the online active learning approach with conventional offline a priori methods.
In a priori learning, the training dataset is predetermined before the inference process begins, without considering the data observed during this process.
We examine three predominant strategies for compiling this dataset:

\begin{itemize}
    \item Grid: the training data forms a structured grid;
    \item LHS: the training data is uniformly distributed and sampled according to a Latin hypercube design~\cite{McKay1979};
    \item Prior: the training data is sampled from the prior distribution of the material parameters.
\end{itemize}

\begin{figure}
    \centering
    \includegraphics[width=0.95\textwidth]{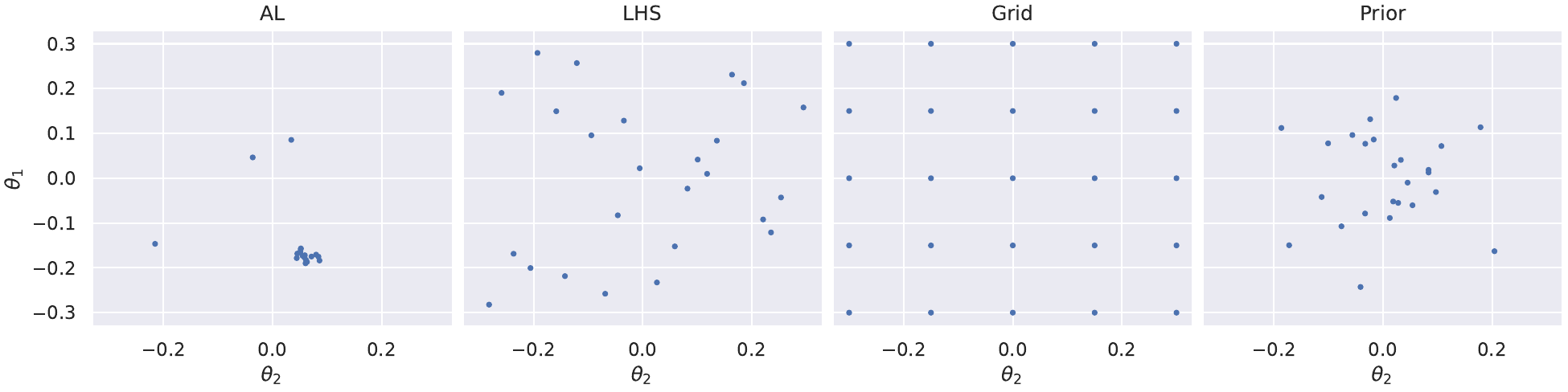}
    \caption{Visualisation of the training data for the 2-dimensional problem. The left subplot shows the training data for the active learning approach, while the remaining subplots show the training data for the a priori learning approaches.}
    \label{fig:train_data_2d}
\end{figure}

\noindent
Despite the simplicity of sampling from the Gaussian prior, both Grid and LHS strategies encounter an implementation challenge due to the prior's infinite support.
Attempting to impose a uniform distribution across this boundless domain inherently results in an improper distribution with zero density, rendering traditional sampling methods impractical.
To address this, we implement a cutoff mechanism to define a finite bounding box in the latent space, enabling meaningful sampling for Grid and LHS methods.
We set this cutoff at three standard deviations from the prior mean in each dimension, prioritising comprehensive coverage over the prior's support.
Although this approach may slightly compromise predictive accuracy within the denser regions of the prior, it significantly enhances model performance in instances where a notable discrepancy exists between the prior and posterior distributions.
The active learning approach of \cref{sec:mcmc_guided_active_learning}, on the other hand, dynamically adjusts the training dataset based on the observed data, thereby adapting to the posterior distribution.
Therefore, it self-determines the amount and location of the training data points.
All four sampling strategies are depicted in \cref{fig:train_data_2d}.

We now want to compare the accuracy of the surrogate models trained with the four different training data collection strategies.
To this end, we consider the range of latent space dimensionalites $d \in \{2, 3, 4, 5\}$ and the associated inverse problems.
For the a priori learning approaches, we set the number of training data points to $N_\mathrm{train} = 5^d$ for an even coverage in all directions.
The GP-hyperparameters are estimated and fixed for the subsequent inference process.
Each setting is repeated 10 times with different ground truths to account for the stochastic nature of the training process and the MCMC\@.
The GP hyperparameter optimisation is performed 20 times with different initialisations.

\begin{figure}
    \centering
    \includegraphics[width=0.98\textwidth]{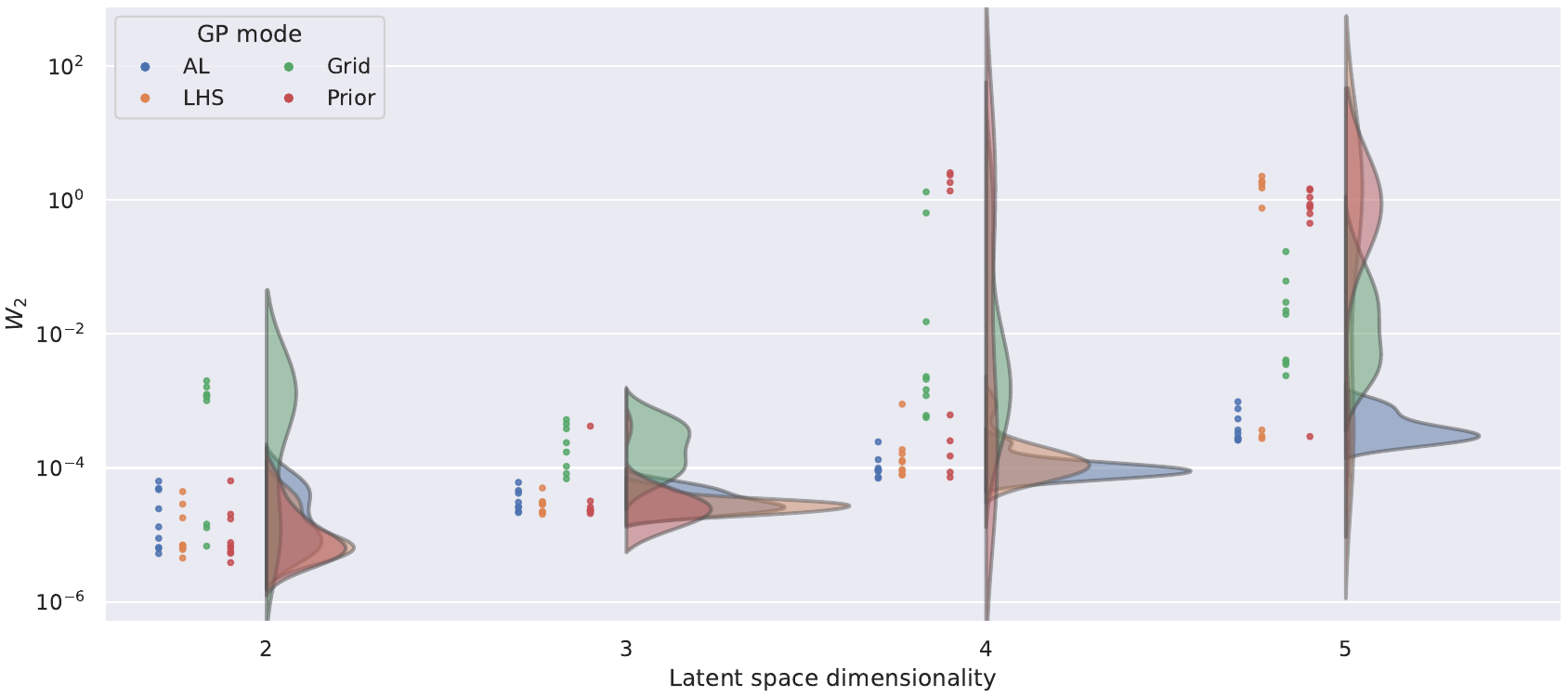}
    \caption{Approximation error of the training data collection strategies for a range of latent space dimensionalities. The half-violins stem from a kernel density estimate of the scattered data in log-space. The offline learning approaches each get $N_{train}=5^d$ training data, while the active learning approach gathers data according to \cref{alg:active_learning}.}
    \label{fig:train_modes_n5_all_raw}
\end{figure}

\cref{fig:train_modes_n5_all_raw} presents the outcomes of our comparative analysis between active learning and a priori training data collection strategies.
We observe that for a latent space with dimensionality of $d=2$ and $d=3$, the performance across all strategies is relatively similar, albeit with grid sampling showing a slight disadvantage.
This parity shifts as we move to higher dimensions ($d=4$ and $d=5$), where the active learning strategy outperforms the a priori methods.
Moreover, a priori learning approaches demonstrate a broader variance in approximation error;
some iterations are nearly as effective as active learning, while others fall markedly short.
This variability is largely attributable to the spatial distribution of training data points with respect to regions of high posterior probability.
In lower-dimensional spaces, the probability of a priori training data aligning with these critical regions remains high.
However, as the dimensionality increases, the likelihood of such fortuitous alignment decreases significantly.
This diminishing probability underlies the observed trends in approximation error:
outcomes are either closely aligned with those of the active learning approach or significantly divergent, with few instances falling in between.
These results underscore the increasing advantage of active learning in higher-dimensional spaces, attributed to its strategic adaptability in navigating the complexities introduced by the curse of dimensionality.

\begin{figure}
    \centering
    \includegraphics[width=0.98\textwidth]{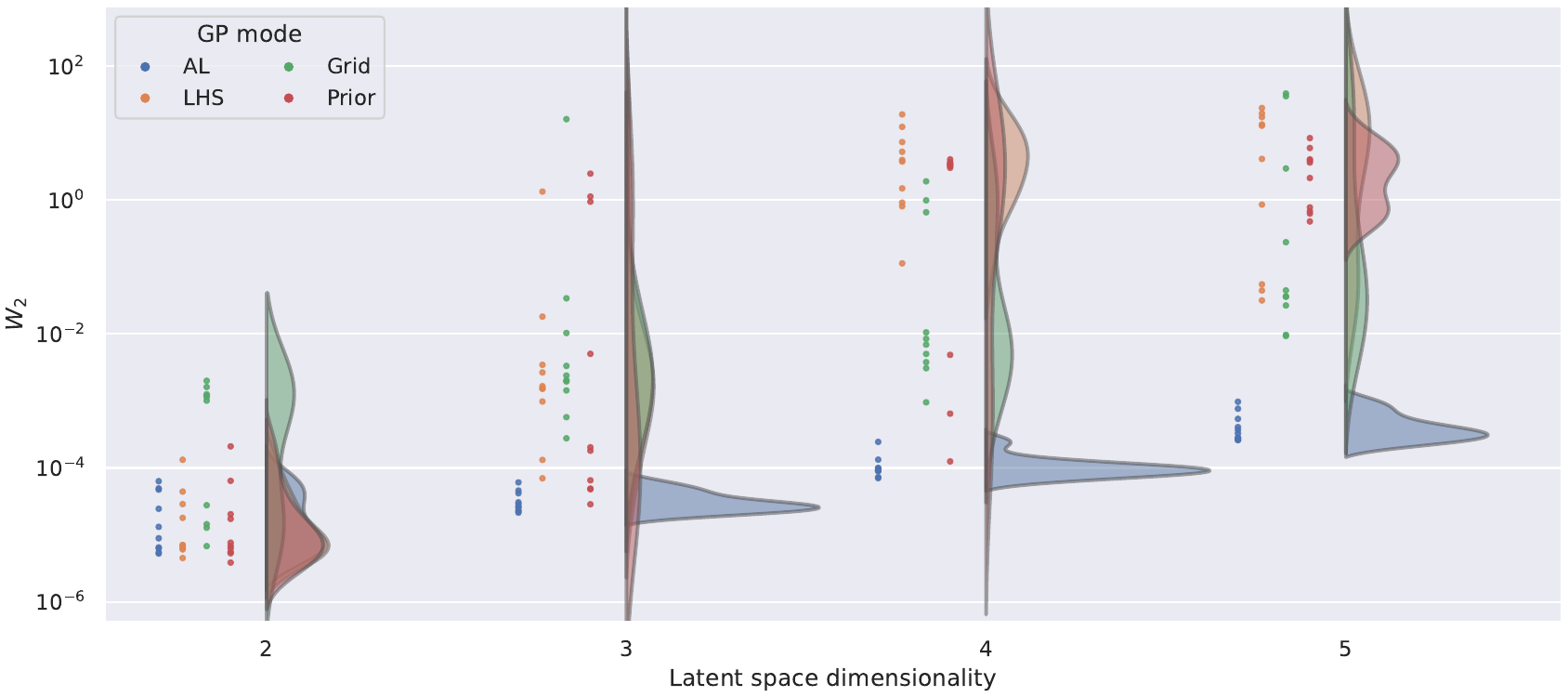}
    \caption{Approximation error of the training data collection strategies for a range of latent space dimensionalities.
    The violins stem from a kernel density estimate of the scatterd data in log-space.
    All strateries get the same amount of training data, as determined by the active learning approach.}
    \label{fig:train_modes_n_al_all_raw}
\end{figure}

The trends observed with the exponential increase in number of training points can be further corroborated by setting the number of training data points to be as similar as possible for all strategies, as determined by the active learning approach.
For this purpose, we set the number of training data points to $N_\mathrm{train} = \lceil \sqrt[d]{N_\mathrm{train,AL}} \rceil ^ d$ for the a priori learning approaches, where $\lceil \cdot \rceil$ denotes smallest integer greater than the argument.
This is slightly favorable for the a priori learning approaches, as $N_\mathrm{train} \geq N_\mathrm{train,AL}$.
Nonetheless, the active learning now clearly outperforms the a priori approaches for all latent space dimensionalities.
Results can be seen in \cref{fig:train_modes_n_al_all_raw}.

\begin{figure}
    \centering
    \subcaptionbox{AL \label{fig:inspection_al_seed-9}}[0.49\textwidth]{
        \includegraphics[width=\linewidth]{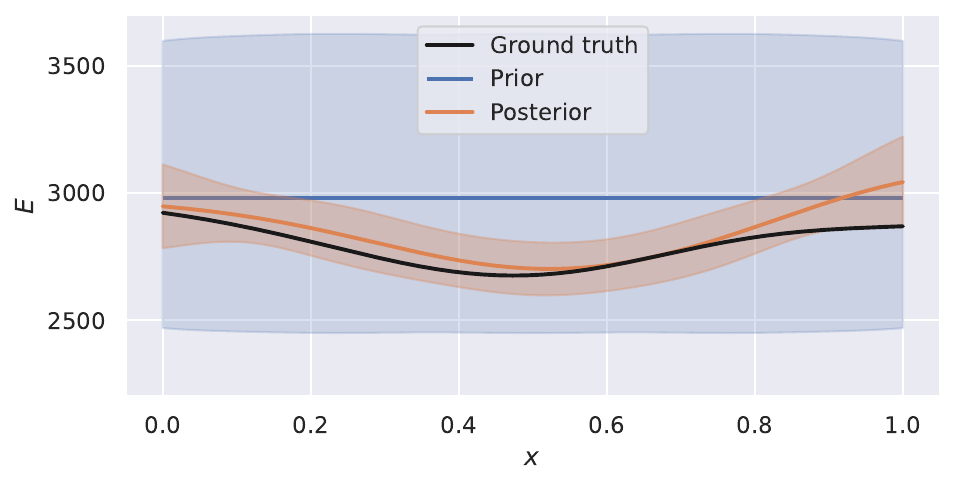}
    }
    \subcaptionbox{LHS \label{fig:inspection_field_lhs_seed-9}}[0.49\textwidth]{
        \includegraphics[width=\linewidth]{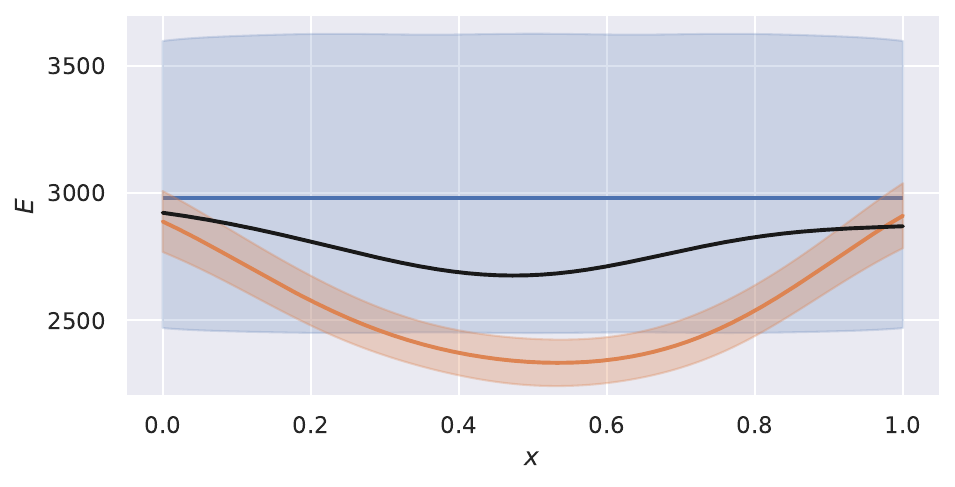}
    }
    \subcaptionbox{AL \label{fig:inspection_pair_plot_cubehelix_al_seed-9}}[0.49\textwidth]{
        \includegraphics[width=\linewidth]{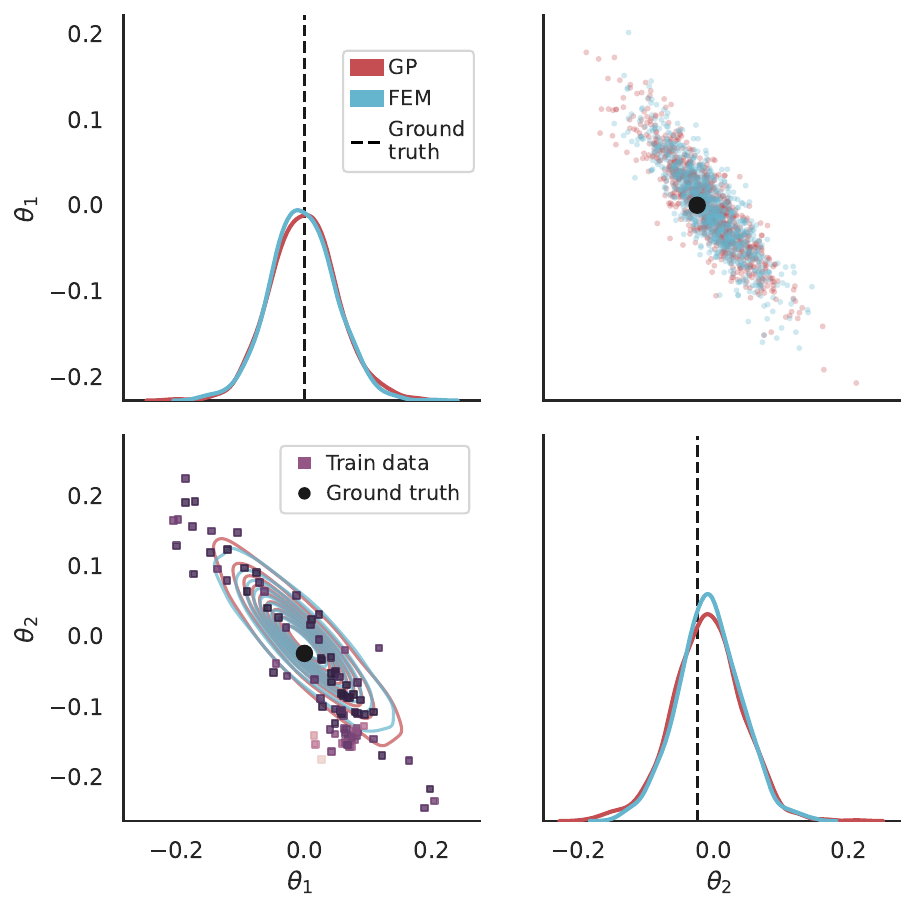}
    }
    \subcaptionbox{LHS \label{fig:inspection_pair_plot_cubehelix_lhs_seed-9}}[0.49\textwidth]{
        \includegraphics[width=\linewidth]{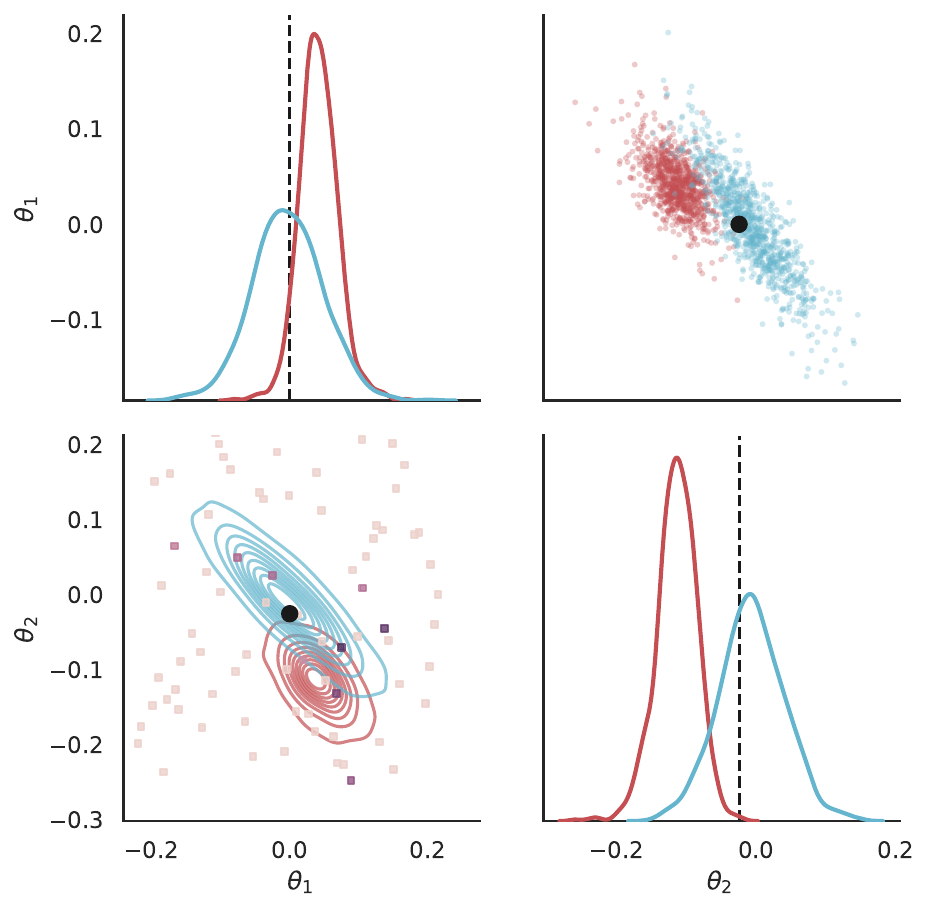}
    }
    \caption{Comparison of the surrogate models trained with the active learning and LHS strategies for a realisation of the 5-dimensional problem.
    The top row shows the true, prior, and posterior fields, while the bottom row shows the pair plots of the first two latent space parameters.
    The training data are color coded---the darker the color, the higher the posterior density.}
    \label{fig:inspection_seed-9}
\end{figure}

As to why the a priori learning approaches perform worse than the active learning approach, it is helpful to inspect the different posteriors resulting from a single run with $d=3$ in \cref{fig:inspection_seed-9}.
The left column displays the results from the AL surrogate model, the right column shows the results from the LHS surrogate model.
The top row shows the prior and posterior random field realisations alongside the ground truth.
The regions close to the boundaries are well represented by the LHS surrogate model in \cref{fig:inspection_field_lhs_seed-9}, whereas the inner part of the domain exposes a significant approximation error compared to the AL solution \cref{fig:inspection_al_seed-9}.
This discrepancy is further highlighted in the pair plots of the first two latent variables in the bottom row (\cref{fig:inspection_pair_plot_cubehelix_al_seed-9,fig:inspection_pair_plot_cubehelix_lhs_seed-9}).
The posterior from the AL surrogate model, depicted in red, is in good agreement with the FEM posterior, shown in cyan.
The training data samples are concentrated in regions of high posterior probability, as indicated by their darker colors.
On the contrary, the LHS surrogate model fails to match the reference posterior, as illustrated by the misalignment of the red and cyan samples in \cref{fig:inspection_pair_plot_cubehelix_lhs_seed-9}.
A closer look at the training data reveals the shortcoming of the LHS strategy:
while the 2-dimensional projection suggest a good coverage of the latent space, the samples are not concentrated in regions of high posterior probability.
Only a handful of points ended up in the region of interest, leading to a poor approximation of the posterior distribution.
This misalignment is naturally more pronounced in higher-dimensional settings, but can already introduce large approximation errors for a latent space dimensionality as low as $d=3$.

 \subsection{Performance and Accuracy Comparison}\label{sec:sampling-efficiency-with-increasing-dimensionality}

\noindent
After exposing the deficiencies of the a priori training data collection strategies, we settle on the active learning approach for the remainder of the experiments.
In this section, we consider the MALA alongside the RWM algorithm, and compare their impact on surrogate model construction, sampling efficiency, and accuracy.
We investigate the performance of the MALA and the RWM algorithm across a range of latent space dimensionalities, from $d=2$ to $d=15$.
Every setting was repeated with 50 different ground-truth realisations (\cref{fig:problem_sequence}) to account for the inherent randomness of the MCMC algorithms.
The GP hyperparameter optimisation is performed 50 times with different initialisations.

\subsubsection{Surrogate Model Construction}\label{sec:surrogate-model-construction}

\noindent
We begin by evaluating the number of training data points across various latent space dimensions and for three distinct reject thresholds $\gamma_{\mathrm{v}}$.
Since the proposal mechanism of the sampling algorithms drives the training data acquisition, the MALA and the RWM algorithm construct different surrogate models.

\begin{figure}
    \captionsetup[subfigure]{aboveskip=-0.1cm, belowskip=0.2cm}
    \centering
    \subcaptionbox{$N_{\mathrm{train}}$ for $\gamma_{\mathrm{v}} = 1.0$ \label{fig:n_data_rej-1.0}}[0.49\textwidth]{
        \includegraphics[width=\linewidth]{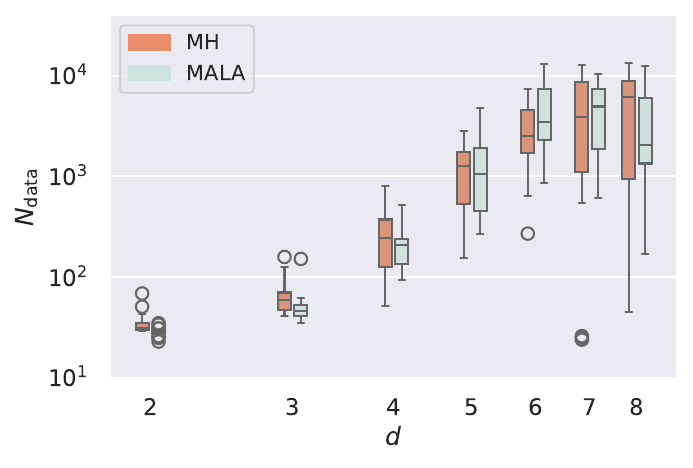}
    }
    \subcaptionbox{$W_2$ for $\gamma_{\mathrm{v}} = 1.0$ \label{fig:wd_rej-1.0}}[0.49\textwidth]{
        \includegraphics[width=\linewidth]{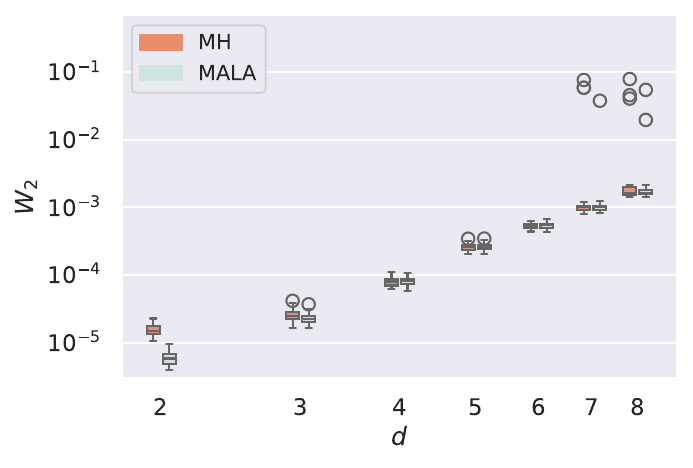}
    }
    \subcaptionbox{$N_{\mathrm{train}}$ for $\gamma_{\mathrm{v}} = 5.0$ \label{fig:n_data_rej-5.0}}[0.49\textwidth]{
        \includegraphics[width=\linewidth]{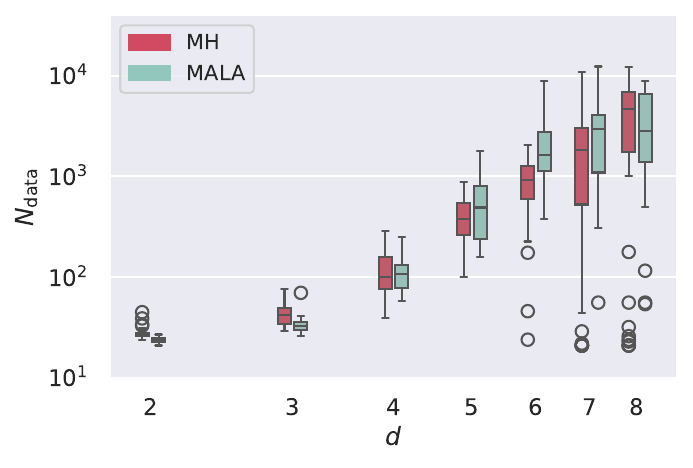}
    }
    \subcaptionbox{$W_2$ for $\gamma_{\mathrm{v}} = 5.0$ \label{fig:wd_rej-5.0}}[0.49\textwidth]{
        \includegraphics[width=\linewidth]{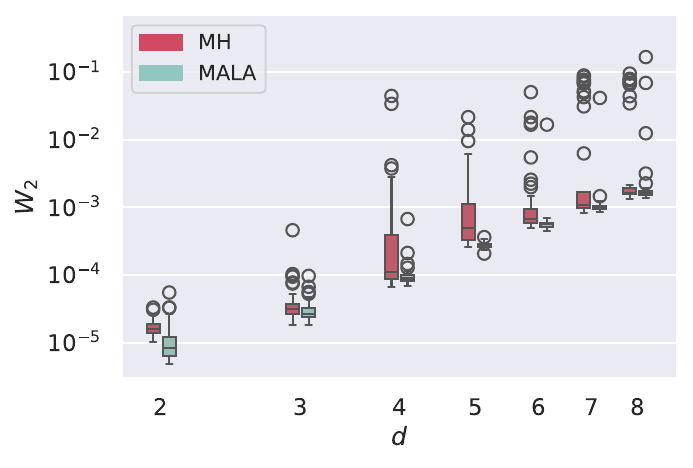}
    }
    \smallskip\\
    \subcaptionbox{$N_{\mathrm{train}}$ for $\gamma_{\mathrm{v}} = 20.0$ \label{fig:n_data_rej-20.0}}[0.49\textwidth]{
        \includegraphics[width=\linewidth]{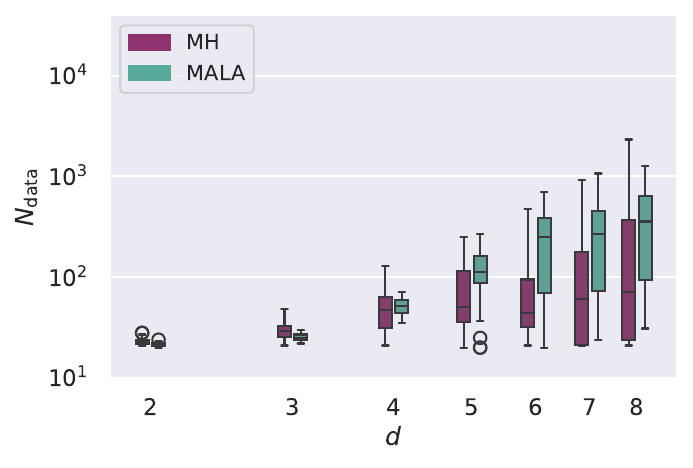}
    }
    \subcaptionbox{$W_2$ for $\gamma_{\mathrm{v}} = 20.0$ \label{fig:wd_rej-20.0}}[0.49\textwidth]{
        \includegraphics[width=\linewidth]{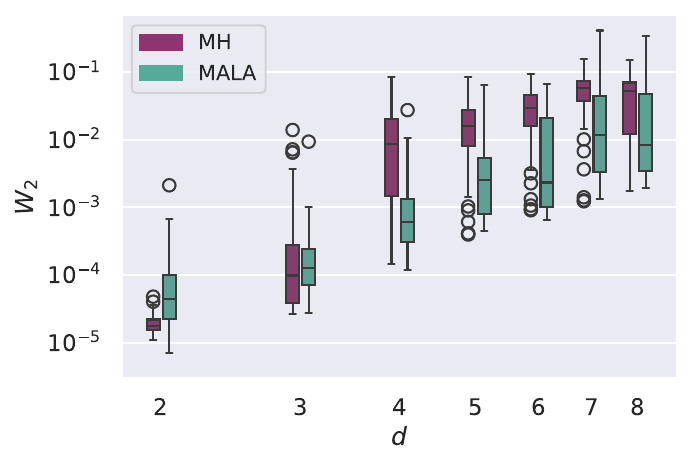}
    }
    \caption{Number of FEM model calls $N_{\mathrm{train}}$ (left column) and Wasserstein 2-distance $W_2$ (right column) for the RWM algorithm and the MALA across a range of latent space dimensionalities $d$ for different reject thresholds $\gamma_{\mathrm{v}}$.}
    \label{fig:surrogate_model_construction}
\end{figure}

The MALA requires fewer training data points than the RWM algorithm across all latent space dimensionalities (\cref{fig:surrogate_model_construction}) under stringent uncertainty requirements.
With the added value of the gradient information, the MALA generates more informative proposals, which are more likely to be accepted.
Therefore, the MALA requires fewer steps to generate sufficient training data for the surrogate model.
The MALA advantage regarding the training data is persistent across all $d$, shown by the constant offset to RWM in log-space.
The slopes of both algorithms, however, are similar, indicating that they are subject to the curse of dimensionality in a similar manner.

The right column of \cref{fig:surrogate_model_construction} details the accuracy of the inferences.
Under strict uncertainty controls ($\gamma_{\mathrm{v}}=1.0$), both RWM and MALA achieve similar levels of accuracy in approximating the posterior distribution.
Outliers below the trend line in \cref{fig:n_data_rej-1.0} correspond to the outliers with high $W_2$ in \cref{fig:wd_rej-1.0}, highlighting the tradeoff between accuracy and efficiency.
These outliers suggest a superior robustness of active learning with the RWM, as they exclusively occurred with the MALA.

With looser uncertainty requirements these outliers become more prevalent (\cref{fig:n_data_rej-5.0,fig:wd_rej-5.0}) and eventually they become the norm (\cref{fig:n_data_rej-20.0,fig:wd_rej-20.0}).
What happens in these MALA runs is that the initial MCMC samples are too far removed from the posterior's high-density regions.
These samples then poorly represent the likelihood's complexity, potentially leading to an overestimated GP length scale and an overly confident surrogate model that ceases to collect more training data.
While the RMW algorithm encounters the same issue, it is less pronounced due to the more exploratory nature of the random walk proposal:
eventually the algorithm moves towards the high-density regions of the posterior by chance, leading to further collection of training data.
The MALA, however, is misled by an erroneous surrogate model and its poor approximation of the gradient and cannot correct itself.

To summarise the findings, the RWM algorithm and the MALA perform similarly both in terms of the number of training data points and the accuracy of the inference when strong requirements are imposed on the uncertainty in the surrogate model.
When the uncertainty tolerance is relaxed, the RWM still produces accurate surrogates, underlining its superior robustness, while the accuracy of the surrogates trained with the MALA deteriorates.

\subsubsection{Adaptation of the Surrogate Model}

\begin{figure}
    \centering
    \subcaptionbox{RWM rate \label{subfig:adaptation_rwm_rate}}[0.48\textwidth]{
        \includegraphics[width=\linewidth]{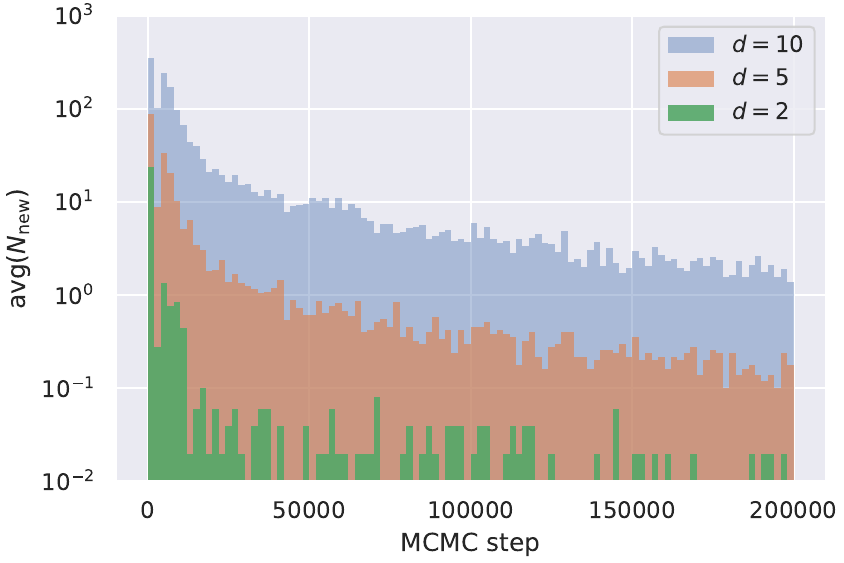}
    }
    \subcaptionbox{MALA rate \label{subfig:adaptation_mala_rate}}[0.48\textwidth]{
        \includegraphics[width=\linewidth]{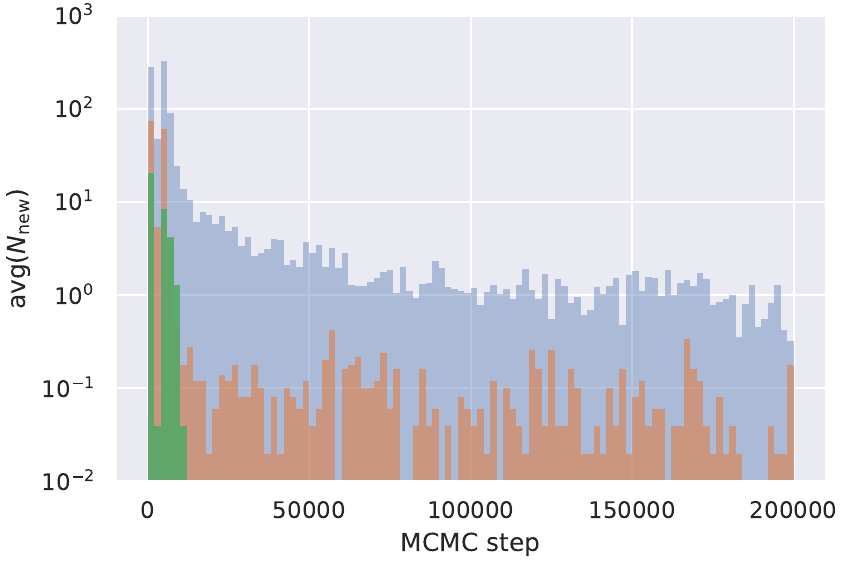}
    }
    \subcaptionbox{RWM cumulative \label{subfig:adaptation_rwm_cum}}[0.48\textwidth]{
        \includegraphics[width=\linewidth]{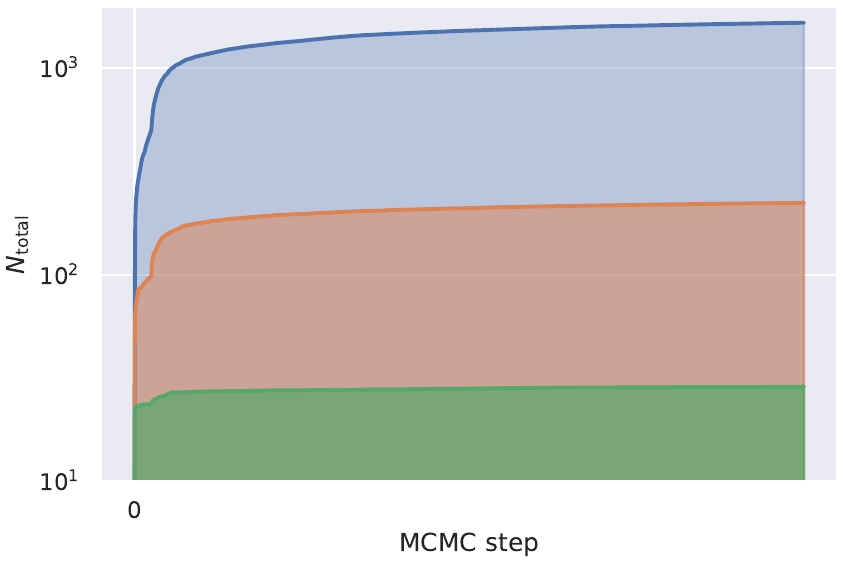}
    }
    \subcaptionbox{MALA cumulative \label{subfig:adaptation_mala_cum}}[0.48\textwidth]{
        \includegraphics[width=\linewidth]{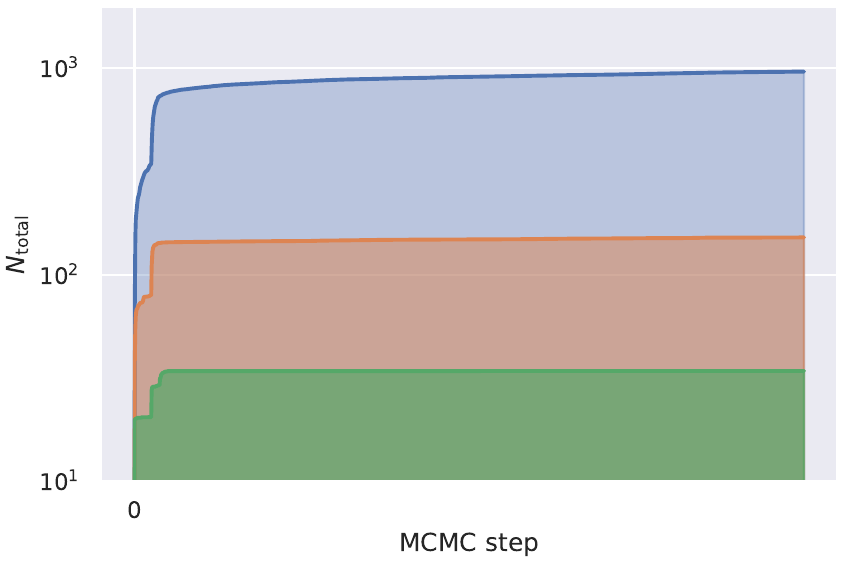}
    }
    \caption{Level of adaptation of the MCMC algorithms RWM in \cref{subfig:adaptation_rwm_rate} and MALA in \cref{subfig:adaptation_mala_rate}.
    Each bar represents the number of new data points collected within intervals of 2000 MCMC steps, averaged over 50 realisations.
    The latent space dimensionality is color coded.}
    \label{fig:adaptation}
\end{figure}

\noindent
While the MCMC algorithms are guaranteed to converge to the distribution given by the surrogate model, the surrogate model itself must also converge to the true likelihood function.
To investigate the level of adaptation of the surrogate model, we analyze the number of new data points collected throughout the MCMC chain.
\cref{fig:adaptation} shows the rate of data points collected per interval of 2000 MCMC steps for the RWM (\cref{subfig:adaptation_rwm_rate}) and MALA (\cref{subfig:adaptation_mala_rate}) algorithms, averaged over 50 MCMC chains each.
The cumulative number of data points collected for both algorithms is shown in \cref{subfig:adaptation_rwm_cum} and \cref{subfig:adaptation_mala_cum}.
Three conclusions can be drawn:
i) the rate of training data collection decreases in all cases,
ii) the rate is higher for the RMW algorithm than for the MALA,
iii) the rate depends on latent space dimensionality---the higher the dimensionality, the more data points are needed at all stages of the chain.
Even though the rate reduces to one new data point per 2000 MCMC steps towards the end of the $d = 10$ case, this dependence suggests a stationary surrogate model becomes more difficult to obtain with increasing latent space dimensionality.

\subsubsection{Sampling Performance Given a Trained Surrogate Model}

\begin{figure}
    \centering
    \includegraphics[width=.55\textwidth]{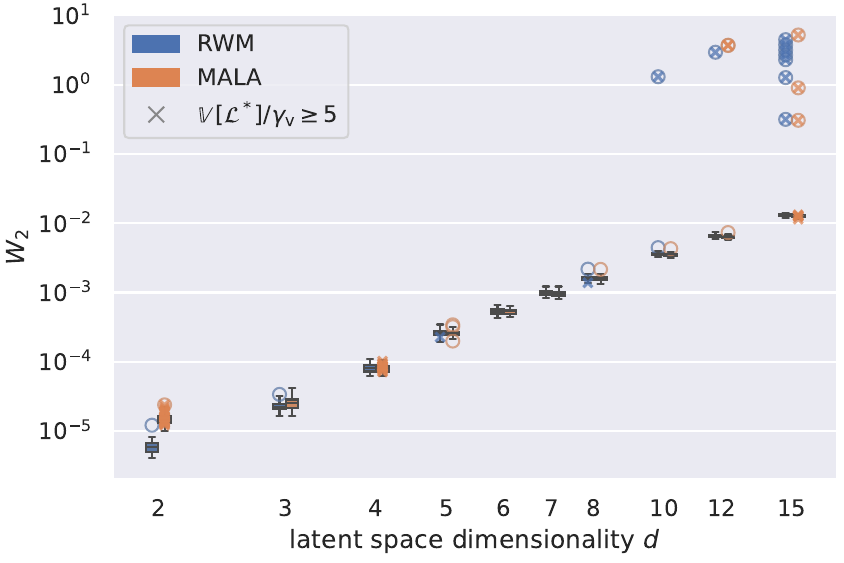}
    \caption{Accuracy of the approximate inference when the RWM algorithm and the MALA use the same surrogate model, across a range of latent space dimensionalities. The crosses in indicate the outliers with predicive variance.}
    \label{fig:sampler_influence}
\end{figure}

\noindent
Finally, we compare the sampling accuracy and efficiency of the RWM algorithm and the MALA when using the same surrogate model.
In this case, we are using the one constructed by the RWM algorithm with $\gamma_{\mathrm{v}} = 1.0$ and prohibit further training data collection.
The numerical experiment is repeated for the same 50 distinct realisations of the inverse problem per latent space dimensionality as in \cref{sec:surrogate-model-construction}.
There is no clear difference in accuracy of the posterior approximation between the MALA and the RWM algorithm for any given dimensionality.
\cref{fig:sampler_influence} shows the Wasserstein 2-distance across a range of latent space dimensionalities.

Starting from $d = 10$, we observe an increasing number of outliers with high $W_2$ values.
These outliers represent chains that have completely diverged from regions with high posterior density.
These chains navigated to largely unexplored regions of the parameter space, as indicated by the large predictive variance encountered during the respective MCMC run.
Runs with at least one occurrence of a maximum predictive variance five times larger than $\gamma_{\mathrm{v}}$ are marked with a cross, which match well with the outliers in \cref{fig:sampler_influence}.
This phenomenon occurs for both algorithms, but is more pronounced for the RWM algorithm.
It underscores the shortcomings of a priori trained surrogate models in high-dimensional settings, as even a model tailored to a given posterior cannot guarantee the accuracy of the inference.

 \section{Conclusions}\label{sec:concusions}

\noindent
In this research, we have examined the integration of surrogate modelling with Markov chain Monte Carlo (MCMC) sampling techniques to improve the efficiency of Bayesian model calibration.
We introduced a simple, scalable one-dimensional bar problem equipped with a non-linear constitutive model and spatially varying stiffness.
The stiffness field is discretised with radial basis functions, the number of which can be adapted to control the dimensionality of the latent space.
This problem serves as our test bed to explore the impacts of various methodological decisions associated with MCMC and surrogate applications.

We employed a Gaussian Process (GP) surrogate model to approximate the log likelihood function and evaluated several traditional a priori training strategies:
Latin hypercube sampling, uniform grid sampling, and sampling from the prior distribution.
Furthermore, we proposed an active learning strategy that adaptively selects training data points based on the MCMC path, leveraging the GP model's inherent uncertainty quantification capabilities.
Our findings underscore the limitations of traditional a priori methods and demonstrate the superiority of active learning in balancing computational costs with analytical accuracy, particularly for problems with higher stochastic dimensionality.

Comparative analysis of the random walk Metropolis (RWM) and Metropolis-adjusted Langevin algorithm (MALA) in terms of surrogate model construction, sampling efficiency, and inference accuracy revealed that both algorithms perform comparably under stringent surrogate model uncertainty requirements.
The increased sampling efficiency of the MALA was beneficial for the training of the surrogate model.
The higher quality of the MALA samples lead to the construction of a surrogate model with similar accuracy to the RWM algorithm, but with fewer FEM model evaluations.
However, under relaxed conditions, accuracy was better maintained with the RWM algorithm than with the MALA, underscoring the robustness of the RWM algorithm.
When a predefined surrogate model was used, the choice of the sampling algorithm did not affect inference accuracy.
However, there is a notable chance for both algorithms to depart from the regions of high surrogate-model certainty and accuracy.
These divergent chains could not return to the high-density regions, even for surrogate models tailored to the posterior distribution, which further emphasised the shortcomings of a priori training.

This study suggest that related works~\cite{Deveney2023, YuanHu2024, Thomas2022, Chen2023} incorrectly identify the MCMC algorithm as a bottleneck in the inference process.
The intensive training data requirements for constructing a surrogate model in high-dimensional scenarios render the MCMC-with-surrogate approach impractical before any significant benefits of advanced sampling algorithms can be realised.
Assuming a forward evaluation of our simulator takes a mere 10 min, the compilation of the training dataset for a 15-dimensional latent space would require roughly 70 days.
Any gains in sampling efficiency from advanced MCMC algorithms would be overshadowed by the prohibitive costs of surrogate model construction.
This finding is at odds with common practice, where engineers often put significant resources into the integration and analysis of advanced MCMC algorithms.
This study clearly suggests prioritising the surrogate model construction over the choice of sampling algorithm.
Further research into multi-fidelity MCMC schemes~\cite{Zhang2018} and nonlinear dimensionality reduction techniques~\cite{Dasgupta2024} could provide viable strategies to mitigate the curse of dimensionality.

\section*{CRediT Authorship Contribution Statement}
\noindent
\textbf{L. Riccius}: Conceptualization, Methodology, Software, Validation, Investigation, Visualization, Writing - Original Draft.
\textbf{I.B.C.M. Rocha}: Conceptualization, Methodology, Supervision, Writing - Review \& Editing.
\textbf{J. Bierkens}: Methodology, Writing - Review \& Editing.
\textbf{H. Kekkonen}: Methodology, Writing - Review \& Editing.
\textbf{F.P. van der Meer}: Conceptualization, Supervision, Writing - Review \& Editing.

\section*{Declaration of Competing Interest}
\noindent
The authors declare that they have no known competing financial interests or personal relationships that could have appeared to influence the work reported in this paper.

\section*{Acknowledgements}
\noindent
This work is supported by the TU Delft AI Labs programme through the SLIMM AI lab.

\section*{Code Availability}
\noindent
The code used in this study is available at \href{https://github.com/SLIMM-Lab/mcmc-with-surrogates.git}{\texttt{github.com/SLIMM-Lab/mcmc-with-surrogates}}.

\newpage
\bibliographystyle{elsarticle-num-names}
\bibliography{export}

\begin{thebibliography}{44}
\expandafter\ifx\csname natexlab\endcsname\relax\def\natexlab#1{#1}\fi
\providecommand{\url}[1]{\texttt{#1}}
\providecommand{\href}[2]{#2}
\providecommand{\path}[1]{#1}
\providecommand{\DOIprefix}{doi:}
\providecommand{\ArXivprefix}{arXiv:}
\providecommand{\URLprefix}{URL: }
\providecommand{\Pubmedprefix}{pmid:}
\providecommand{\doi}[1]{\href{http://dx.doi.org/#1}{\path{#1}}}
\providecommand{\Pubmed}[1]{\href{pmid:#1}{\path{#1}}}
\providecommand{\bibinfo}[2]{#2}
\ifx\xfnm\relax \def\xfnm[#1]{\unskip,\space#1}\fi
\bibitem[{Dobrilla et~al.(2023)Dobrilla, Lunardelli, Nikolić, Lowke, and
  Rosić}]{Dobrilla2023}
\bibinfo{author}{S.~Dobrilla}, \bibinfo{author}{M.~Lunardelli},
  \bibinfo{author}{M.~Nikolić}, \bibinfo{author}{D.~Lowke},
  \bibinfo{author}{B.~Rosić},
\newblock \bibinfo{title}{Bayesian inference of mesoscale mechanical properties
  of mortar using experimental data from a double shear test},
\newblock \bibinfo{journal}{Computer Methods in Applied Mechanics and
  Engineering} \bibinfo{volume}{409} (\bibinfo{year}{2023})
  \bibinfo{pages}{115964}. \URLprefix
  \url{https://linkinghub.elsevier.com/retrieve/pii/S0045782523000877}.
\bibitem[{Marsili et~al.(2017)Marsili, Croce, Friedman, Formichi, and
  Landi}]{Marsili2017}
\bibinfo{author}{F.~Marsili}, \bibinfo{author}{P.~Croce},
  \bibinfo{author}{N.~Friedman}, \bibinfo{author}{P.~Formichi},
  \bibinfo{author}{F.~Landi},
\newblock \bibinfo{title}{Seismic reliability assessment of a concrete water
  tank based on the bayesian updating of the finite element model},
\newblock \bibinfo{journal}{ASCE-ASME Journal of Risk and Uncertainty in
  Engineering Systems, Part B: Mechanical Engineering} \bibinfo{volume}{3}
  (\bibinfo{year}{2017}). \URLprefix
  \url{https://asmedigitalcollection.asme.org/risk/article/3/2/021004/369959/Seismic-Reliability-Assessment-of-a-Concrete-Water}.
\bibitem[{Mengersen and Tweedie(1996)}]{Mengersen1996}
\bibinfo{author}{K.~L. Mengersen}, \bibinfo{author}{R.~L. Tweedie},
\newblock \bibinfo{title}{Rates of convergence of the hastings and metropolis
  algorithms},
\newblock \bibinfo{journal}{The Annals of Statistics} \bibinfo{volume}{24}
  (\bibinfo{year}{1996}) \bibinfo{pages}{101--121}. \URLprefix
  \url{https://projecteuclid.org/journals/annals-of-statistics/volume-24/issue-1/Rates-of-convergence-of-the-Hastings-and-Metropolis-algorithms/10.1214/aos/1033066201.full}.
\bibitem[{Roberts and Rosenthal(2001)}]{Roberts2001}
\bibinfo{author}{G.~O. Roberts}, \bibinfo{author}{J.~S. Rosenthal},
\newblock \bibinfo{title}{Optimal scaling for various metropolis-hastings
  algorithms},
\newblock \bibinfo{journal}{https://doi.org/10.1214/ss/1015346320}
  \bibinfo{volume}{16} (\bibinfo{year}{2001}) \bibinfo{pages}{351--367}.
  \URLprefix
  \url{https://projecteuclid.org/journals/statistical-science/volume-16/issue-4/Optimal-scaling-for-various-Metropolis-Hastings-algorithms/10.1214/ss/1015346320.full
  https://projecteuclid.org/journals/statistical-science/volume-16/issue-4/Optimal-scaling-for-various-Metropolis-Hastings-algorithms/10.1214/ss/1015346320.short}.
\bibitem[{Roberts and Stramer(2002)}]{Roberts2002}
\bibinfo{author}{G.~O. Roberts}, \bibinfo{author}{O.~Stramer},
\newblock \bibinfo{title}{Langevin diffusions and metropolis-hastings
  algorithms},
\newblock \bibinfo{journal}{Methodology And Computing In Applied Probability
  2002 4:4} \bibinfo{volume}{4} (\bibinfo{year}{2002})
  \bibinfo{pages}{337--357}. \URLprefix
  \url{https://link.springer.com/article/10.1023/A:1023562417138}.
\bibitem[{Duane et~al.(1987)Duane, Kennedy, Pendleton, and Roweth}]{Duane1987}
\bibinfo{author}{S.~Duane}, \bibinfo{author}{A.~Kennedy},
  \bibinfo{author}{B.~J. Pendleton}, \bibinfo{author}{D.~Roweth},
\newblock \bibinfo{title}{Hybrid monte carlo},
\newblock \bibinfo{journal}{Physics Letters B} \bibinfo{volume}{195}
  (\bibinfo{year}{1987}) \bibinfo{pages}{216--222}. \URLprefix
  \url{https://linkinghub.elsevier.com/retrieve/pii/037026938791197X}.
\bibitem[{Girolami and Calderhead(2011)}]{Girolami2011}
\bibinfo{author}{M.~Girolami}, \bibinfo{author}{B.~Calderhead},
\newblock \bibinfo{title}{Riemann manifold langevin and hamiltonian monte carlo
  methods},
\newblock \bibinfo{journal}{Journal of the Royal Statistical Society. Series B:
  Statistical Methodology} \bibinfo{volume}{73} (\bibinfo{year}{2011})
  \bibinfo{pages}{123--214}. \URLprefix \url{www.ucl.ac.uk/statistics/}.
\bibitem[{Bierkens et~al.(2019)Bierkens, Fearnhead, and Roberts}]{Bierkens2019}
\bibinfo{author}{J.~Bierkens}, \bibinfo{author}{P.~Fearnhead},
  \bibinfo{author}{G.~Roberts},
\newblock \bibinfo{title}{The zig-zag process and super-efficient sampling for
  bayesian analysis of big data},
\newblock \bibinfo{journal}{The Annals of Statistics} \bibinfo{volume}{47}
  (\bibinfo{year}{2019}) \bibinfo{pages}{1288--1320}. \URLprefix
  \url{https://projecteuclid.org/journals/annals-of-statistics/volume-47/issue-3/The-Zig-Zag-process-and-super-efficient-sampling-for-Bayesian/10.1214/18-AOS1715.full}.
\bibitem[{Chong and Lam(2017)}]{Chong2017}
\bibinfo{author}{A.~Chong}, \bibinfo{author}{K.~P. Lam},
\newblock \bibinfo{title}{A comparison of mcmc algorithms for the bayesian
  calibration of building energy models for building simulation 2017
  conference},
\newblock in: \bibinfo{booktitle}{Building Simulation Conference Proceedings},
  volume~\bibinfo{volume}{2}, \bibinfo{year}{2017}, pp.
  \bibinfo{pages}{582--591}. \URLprefix
  \url{https://doi.org/10.26868/25222708.2017.336}.
\bibitem[{Goodman and Weare(2010)}]{Goodman2010}
\bibinfo{author}{J.~Goodman}, \bibinfo{author}{J.~Weare},
\newblock \bibinfo{title}{Ensemble samplers with affine invariance},
\newblock \bibinfo{journal}{Communications in Applied Mathematics and
  Computational Science} \bibinfo{volume}{5} (\bibinfo{year}{2010})
  \bibinfo{pages}{65--80}.
\bibitem[{Gelman et~al.(1997)Gelman, Gilks, and Roberts}]{Gelman1997}
\bibinfo{author}{A.~Gelman}, \bibinfo{author}{W.~R. Gilks},
  \bibinfo{author}{G.~O. Roberts},
\newblock \bibinfo{title}{Weak convergence and optimal scaling of random walk
  metropolis algorithms},
\newblock \bibinfo{journal}{The Annals of Applied Probability}
  \bibinfo{volume}{7} (\bibinfo{year}{1997}) \bibinfo{pages}{110--120}.
  \URLprefix
  \url{https://projecteuclid.org/journals/annals-of-applied-probability/volume-7/issue-1/Weak-convergence-and-optimal-scaling-of-random-walk-Metropolis-algorithms/10.1214/aoap/1034625254.full}.
\bibitem[{Roberts and Rosenthal(1998)}]{Roberts1998}
\bibinfo{author}{G.~O. Roberts}, \bibinfo{author}{J.~S. Rosenthal},
\newblock \bibinfo{title}{Optimal scaling of discrete approximations to
  langevin diffusions},
\newblock \bibinfo{journal}{Journal of the Royal Statistical Society Series B:
  Statistical Methodology} \bibinfo{volume}{60} (\bibinfo{year}{1998})
  \bibinfo{pages}{255--268}. \URLprefix
  \url{https://academic.oup.com/jrsssb/article/60/1/255/7083121}.
\bibitem[{Ching and Chen(2007)}]{Ching2007}
\bibinfo{author}{J.~Ching}, \bibinfo{author}{Y.-C. Chen},
\newblock \bibinfo{title}{Transitional markov chain monte carlo method for
  bayesian model updating, model class selection, and model averaging},
\newblock \bibinfo{journal}{Journal of Engineering Mechanics}
  \bibinfo{volume}{133} (\bibinfo{year}{2007}) \bibinfo{pages}{816--832}.
\bibitem[{Lye et~al.(2022)Lye, Cicirello, and Patelli}]{Lye2022}
\bibinfo{author}{A.~Lye}, \bibinfo{author}{A.~Cicirello},
  \bibinfo{author}{E.~Patelli},
\newblock \bibinfo{title}{An efficient and robust sampler for bayesian
  inference: Transitional ensemble markov chain monte carlo},
\newblock \bibinfo{journal}{Mechanical Systems and Signal Processing}
  \bibinfo{volume}{167} (\bibinfo{year}{2022}) \bibinfo{pages}{108471}.
  \URLprefix \url{https://doi.org/10.1016/j.ymssp.2021.108471}.
\bibitem[{Straub and Papaioannou(2015)}]{Straub2015}
\bibinfo{author}{D.~Straub}, \bibinfo{author}{I.~Papaioannou},
\newblock \bibinfo{title}{Bayesian updating with structural reliability
  methods},
\newblock \bibinfo{journal}{Journal of Engineering Mechanics}
  \bibinfo{volume}{141} (\bibinfo{year}{2015}) \bibinfo{pages}{04014134}.
  \URLprefix \url{https://doi.org/10.1061/(ASCE)EM.1943-7889.0000839}.
\bibitem[{Hu et~al.(2024)Hu, Abuseada, Alghfeli, Holdheim, and
  Fisher}]{YuanHu2024}
\bibinfo{author}{Y.~Hu}, \bibinfo{author}{M.~Abuseada},
  \bibinfo{author}{A.~Alghfeli}, \bibinfo{author}{S.~Holdheim},
  \bibinfo{author}{T.~S. Fisher},
\newblock \bibinfo{title}{Surrogate-accelerated bayesian framework for
  high-temperature thermal diffusivity characterization},
\newblock \bibinfo{journal}{Computer Methods in Applied Mechanics and
  Engineering} \bibinfo{volume}{418} (\bibinfo{year}{2024})
  \bibinfo{pages}{116459}. \URLprefix
  \url{https://doi.org/10.1016/j.cma.2023.116459}.
\bibitem[{Thomas et~al.(2022)Thomas, Barocio, Bilionis, and Pipes}]{Thomas2022}
\bibinfo{author}{A.~J. Thomas}, \bibinfo{author}{E.~Barocio},
  \bibinfo{author}{I.~Bilionis}, \bibinfo{author}{R.~B. Pipes},
\newblock \bibinfo{title}{Bayesian inference of fiber orientation and polymer
  properties in short fiber-reinforced polymer composites}
  (\bibinfo{year}{2022}). \URLprefix \url{https://arxiv.org/abs/2202.12881v1
  http://arxiv.org/abs/2202.12881}.
\bibitem[{Wu et~al.(2020)Wu, Zulueta, Major, Arriaga, and Noels}]{Wu2020a}
\bibinfo{author}{L.~Wu}, \bibinfo{author}{K.~Zulueta},
  \bibinfo{author}{Z.~Major}, \bibinfo{author}{A.~Arriaga},
  \bibinfo{author}{L.~Noels},
\newblock \bibinfo{title}{Bayesian inference of non-linear multiscale model
  parameters accelerated by a deep neural network},
\newblock \bibinfo{journal}{Computer Methods in Applied Mechanics and
  Engineering} \bibinfo{volume}{360} (\bibinfo{year}{2020}). \URLprefix
  \url{www.sciencedirect.comwww.elsevier.com/locate/cma}, n.
\bibitem[{Deveney et~al.(2023)Deveney, Mueller, and Shardlow}]{Deveney2023}
\bibinfo{author}{T.~Deveney}, \bibinfo{author}{E.~H. Mueller},
  \bibinfo{author}{T.~Shardlow},
\newblock \bibinfo{title}{Deep surrogate accelerated delayed-acceptance
  hamiltonian monte carlo for bayesian inference of spatio-temporal heat fluxes
  in rotating disc systems},
\newblock \bibinfo{journal}{SIAM/ASA Journal on Uncertainty Quantification}
  \bibinfo{volume}{11} (\bibinfo{year}{2023}) \bibinfo{pages}{970--995}.
  \URLprefix \url{https://epubs.siam.org/doi/10.1137/22M1513113}.
\bibitem[{Chen et~al.(2023)Chen, Jin, Li, Qi, and Cai}]{Chen2023}
\bibinfo{author}{Z.~Chen}, \bibinfo{author}{P.~Jin}, \bibinfo{author}{R.~Li},
  \bibinfo{author}{Y.~Qi}, \bibinfo{author}{G.~Cai},
\newblock \bibinfo{title}{Parameter identification of elastoplastic model for
  cucrzr alloy by the neural network‐aided bayesian inference},
\newblock \bibinfo{journal}{Fatigue \& Fracture of Engineering Materials \&
  Structures} \bibinfo{volume}{46} (\bibinfo{year}{2023})
  \bibinfo{pages}{2319--2337}. \URLprefix
  \url{https://onlinelibrary.wiley.com/doi/10.1111/ffe.14000}.
\bibitem[{Drovandi et~al.(2018)Drovandi, Moores, and Boys}]{Drovandi2018}
\bibinfo{author}{C.~C. Drovandi}, \bibinfo{author}{M.~T. Moores},
  \bibinfo{author}{R.~J. Boys},
\newblock \bibinfo{title}{Accelerating pseudo-marginal mcmc using gaussian
  processes},
\newblock \bibinfo{journal}{Computational Statistics \& Data Analysis}
  \bibinfo{volume}{118} (\bibinfo{year}{2018}) \bibinfo{pages}{1--17}.
\bibitem[{del Val et~al.(2022)del Val, Maître, Magin, Chazot, and
  Congedo}]{DelVal2022}
\bibinfo{author}{A.~del Val}, \bibinfo{author}{O.~P.~L. Maître},
  \bibinfo{author}{T.~E. Magin}, \bibinfo{author}{O.~Chazot},
  \bibinfo{author}{P.~M. Congedo},
\newblock \bibinfo{title}{A surrogate-based optimal likelihood function for the
  bayesian calibration of catalytic recombination in atmospheric entry
  protection materials},
\newblock \bibinfo{journal}{Applied Mathematical Modelling}
  \bibinfo{volume}{101} (\bibinfo{year}{2022}) \bibinfo{pages}{791--810}.
\bibitem[{Zhang et~al.(2018)Zhang, Man, Lin, Wu, and Zeng}]{Zhang2018}
\bibinfo{author}{J.~Zhang}, \bibinfo{author}{J.~Man}, \bibinfo{author}{G.~Lin},
  \bibinfo{author}{L.~Wu}, \bibinfo{author}{L.~Zeng},
\newblock \bibinfo{title}{Inverse modeling of hydrologic systems with adaptive
  multifidelity markov chain monte carlo simulations},
\newblock \bibinfo{journal}{Water Resources Research} \bibinfo{volume}{54}
  (\bibinfo{year}{2018}) \bibinfo{pages}{4867--4886}. \URLprefix
  \url{https://onlinelibrary.wiley.com/doi/full/10.1029/2018WR022658
  https://onlinelibrary.wiley.com/doi/abs/10.1029/2018WR022658
  https://agupubs.onlinelibrary.wiley.com/doi/10.1029/2018WR022658}.
\bibitem[{Dasgupta et~al.(2024)Dasgupta, Patel, Ray, Johnson, and
  Oberai}]{Dasgupta2024}
\bibinfo{author}{A.~Dasgupta}, \bibinfo{author}{D.~V. Patel},
  \bibinfo{author}{D.~Ray}, \bibinfo{author}{E.~A. Johnson},
  \bibinfo{author}{A.~A. Oberai},
\newblock \bibinfo{title}{A dimension-reduced variational approach for solving
  physics-based inverse problems using generative adversarial network priors
  and normalizing flows},
\newblock \bibinfo{journal}{Computer Methods in Applied Mechanics and
  Engineering} \bibinfo{volume}{420} (\bibinfo{year}{2024})
  \bibinfo{pages}{116682}. \URLprefix
  \url{https://linkinghub.elsevier.com/retrieve/pii/S0045782523008058}.
\bibitem[{Kandasamy et~al.(2015)Kandasamy, Schneider, and
  Póczos}]{Kandasamy2015}
\bibinfo{author}{K.~Kandasamy}, \bibinfo{author}{J.~Schneider},
  \bibinfo{author}{B.~Póczos},
\newblock \bibinfo{title}{Bayesian active learning for posterior estimation},
\newblock in: \bibinfo{booktitle}{IJCAI International Joint Conference on
  Artificial Intelligence}, volume \bibinfo{volume}{2015-Janua},
  \bibinfo{year}{2015}, pp. \bibinfo{pages}{3605--3611}.
\bibitem[{Dinkel et~al.(2024)Dinkel, Geitner, Rei, Nitzler, and
  Wall}]{Dinkel2024}
\bibinfo{author}{M.~Dinkel}, \bibinfo{author}{C.~M. Geitner},
  \bibinfo{author}{G.~R. Rei}, \bibinfo{author}{J.~Nitzler},
  \bibinfo{author}{W.~A. Wall},
\newblock \bibinfo{title}{Solving bayesian inverse problems with expensive
  likelihoods using constrained gaussian processes and active learning},
\newblock \bibinfo{journal}{Inverse Problems} \bibinfo{volume}{40}
  (\bibinfo{year}{2024}) \bibinfo{pages}{095008}. \URLprefix
  \url{https://iopscience.iop.org/article/10.1088/1361-6420/ad5eb4}.
\bibitem[{Hou et~al.(2021)Hou, Hassan, and Wang}]{Hou2021}
\bibinfo{author}{D.~Hou}, \bibinfo{author}{I.~Hassan},
  \bibinfo{author}{L.~Wang},
\newblock \bibinfo{title}{Review on building energy model calibration by
  bayesian inference},
\newblock \bibinfo{journal}{Renewable and Sustainable Energy Reviews}
  \bibinfo{volume}{143} (\bibinfo{year}{2021}) \bibinfo{pages}{110930}.
  \URLprefix
  \url{https://linkinghub.elsevier.com/retrieve/pii/S1364032121002239}.
\bibitem[{Rappel et~al.(2019)Rappel, Beex, Noels, and Bordas}]{Rappel2019}
\bibinfo{author}{H.~Rappel}, \bibinfo{author}{L.~A. Beex},
  \bibinfo{author}{L.~Noels}, \bibinfo{author}{S.~P. Bordas},
\newblock \bibinfo{title}{Identifying elastoplastic parameters with bayes’
  theorem considering output error, input error and model uncertainty},
\newblock \bibinfo{journal}{Probabilistic Engineering Mechanics}
  \bibinfo{volume}{55} (\bibinfo{year}{2019}) \bibinfo{pages}{28--41}.
\bibitem[{Rappel et~al.(2020)Rappel, Beex, Hale, Noels, and
  Bordas}]{Rappel2020}
\bibinfo{author}{H.~Rappel}, \bibinfo{author}{L.~A. Beex},
  \bibinfo{author}{J.~S. Hale}, \bibinfo{author}{L.~Noels},
  \bibinfo{author}{S.~P. Bordas},
\newblock \bibinfo{title}{A tutorial on bayesian inference to identify material
  parameters in solid mechanics},
\newblock \bibinfo{journal}{Archives of Computational Methods in Engineering}
  \bibinfo{volume}{27} (\bibinfo{year}{2020}) \bibinfo{pages}{361--385}.
\bibitem[{Sudret and Kiureghian(2000)}]{Sudret2000}
\bibinfo{author}{B.~Sudret}, \bibinfo{author}{A.~D. Kiureghian},
  \bibinfo{title}{Stochastic finite element methods and reliability: a
  state-of-the-art report}, \bibinfo{year}{2000}. \URLprefix
  \url{https://books.google.nl/books?id=PNNztgAACAAJ}.
\bibitem[{Marzouk and Najm(2009)}]{Marzouk2009}
\bibinfo{author}{Y.~M. Marzouk}, \bibinfo{author}{H.~N. Najm},
\newblock \bibinfo{title}{Dimensionality reduction and polynomial chaos
  acceleration of bayesian inference in inverse problems},
\newblock \bibinfo{journal}{Journal of Computational Physics}
  \bibinfo{volume}{228} (\bibinfo{year}{2009}) \bibinfo{pages}{1862--1902}.
  \URLprefix
  \url{https://linkinghub.elsevier.com/retrieve/pii/S0021999108006062}.
\bibitem[{Nouy and Maître(2009)}]{Nouy2009}
\bibinfo{author}{A.~Nouy}, \bibinfo{author}{O.~P.~L. Maître},
\newblock \bibinfo{title}{Generalized spectral decomposition for stochastic
  nonlinear problems},
\newblock \bibinfo{journal}{Journal of Computational Physics}
  \bibinfo{volume}{228} (\bibinfo{year}{2009}) \bibinfo{pages}{202--235}.
  \URLprefix
  \url{https://linkinghub.elsevier.com/retrieve/pii/S0021999108004737}.
\bibitem[{Vigliotti et~al.(2018)Vigliotti, Csányi, and
  Deshpande}]{Vigliotti2018}
\bibinfo{author}{A.~Vigliotti}, \bibinfo{author}{G.~Csányi},
  \bibinfo{author}{V.~S. Deshpande},
\newblock \bibinfo{title}{Bayesian inference of the spatial distributions of
  material properties},
\newblock \bibinfo{journal}{Journal of the Mechanics and Physics of Solids}
  \bibinfo{volume}{118} (\bibinfo{year}{2018}) \bibinfo{pages}{74--97}.
\bibitem[{Li and Kiureghian(1993)}]{Li1993}
\bibinfo{author}{C.~Li}, \bibinfo{author}{A.~D. Kiureghian},
\newblock \bibinfo{title}{Optimal discretization of random fields},
\newblock \bibinfo{journal}{Journal of Engineering Mechanics}
  \bibinfo{volume}{119} (\bibinfo{year}{1993}) \bibinfo{pages}{1136--1154}.
  \URLprefix
  \url{https://ascelibrary.org/doi/abs/10.1061/%28ASCE%290733-9399%281993%29119%3A6%281136%29
  https://ascelibrary.org/doi/10.1061/%28ASCE%290733-9399%281993%29119%3A6%281136%29}.
\bibitem[{Ghanem and Spanos(1991)}]{Ghanem1991}
\bibinfo{author}{R.~G. Ghanem}, \bibinfo{author}{P.~D. Spanos},
  \bibinfo{title}{Stochastic Finite Elements: A Spectral Approach},
  \bibinfo{publisher}{Springer New York}, \bibinfo{year}{1991}. \URLprefix
  \url{http://link.springer.com/10.1007/978-1-4612-3094-6}.
\bibitem[{Rasmussen and Williams(2006)}]{Rasmussen2006}
\bibinfo{author}{C.~E. Rasmussen}, \bibinfo{author}{C.~K.~I. Williams},
  \bibinfo{title}{Gaussian processes for machine learning.},
  \bibinfo{publisher}{MIT Press}, \bibinfo{year}{2006}. \URLprefix
  \url{www.GaussianProcess.org/gpml}.
\bibitem[{Mackay(1998)}]{Mackay1998}
\bibinfo{author}{D.~J.~C. Mackay}, \bibinfo{title}{Introduction to Gaussian
  Processes}, \bibinfo{publisher}{Barber and Williams}, \bibinfo{year}{1998}.
  \URLprefix \url{http://www.cs.toronto.edu/~radford/.}
\bibitem[{Stuart(2010)}]{Stuart2010}
\bibinfo{author}{A.~M. Stuart}, \bibinfo{title}{Inverse problems: A bayesian
  perspective}, \bibinfo{year}{2010}. \URLprefix
  \url{https://doi.org/10.1017/S0962492910000061}.
\bibitem[{Simoen et~al.(2013)Simoen, Papadimitriou, and Lombaert}]{Simoen2013}
\bibinfo{author}{E.~Simoen}, \bibinfo{author}{C.~Papadimitriou},
  \bibinfo{author}{G.~Lombaert},
\newblock \bibinfo{title}{On prediction error correlation in bayesian model
  updating},
\newblock \bibinfo{journal}{Journal of Sound and Vibration}
  \bibinfo{volume}{332} (\bibinfo{year}{2013}) \bibinfo{pages}{4136--4152}.
  \URLprefix
  \url{https://linkinghub.elsevier.com/retrieve/pii/S0022460X13002514}.
\bibitem[{Roberts and Rosenthal(2007)}]{Roberts2007}
\bibinfo{author}{G.~O. Roberts}, \bibinfo{author}{J.~S. Rosenthal},
\newblock \bibinfo{title}{Coupling and ergodicity of adaptive markov chain
  monte carlo algorithms},
\newblock \bibinfo{journal}{Journal of Applied Probability}
  \bibinfo{volume}{44} (\bibinfo{year}{2007}) \bibinfo{pages}{458--475}.
  \URLprefix
  \url{https://www.cambridge.org/core/product/identifier/S0021900200117954/type/journal_article}.
\bibitem[{Fletcher(2000)}]{Fletcher2000}
\bibinfo{author}{R.~Fletcher},
\newblock \bibinfo{title}{Practical methods of optimization},
\newblock \bibinfo{journal}{Practical Methods of Optimization}
  (\bibinfo{year}{2000}). \URLprefix
  \url{https://onlinelibrary.wiley.com/doi/book/10.1002/9781118723203}.
\bibitem[{Rocha et~al.(2021)Rocha, Kerfriden, and van~der Meer}]{Rocha2021}
\bibinfo{author}{I.~B. Rocha}, \bibinfo{author}{P.~Kerfriden},
  \bibinfo{author}{F.~P. van~der Meer},
\newblock \bibinfo{title}{On-the-fly construction of surrogate constitutive
  models for concurrent multiscale mechanical analysis through probabilistic
  machine learning},
\newblock \bibinfo{journal}{Journal of Computational Physics: X}
  \bibinfo{volume}{9} (\bibinfo{year}{2021}) \bibinfo{pages}{100083}.
\bibitem[{Melro et~al.(2013)Melro, Camanho, Pires, and Pinho}]{Melro2013}
\bibinfo{author}{A.~R. Melro}, \bibinfo{author}{P.~P. Camanho},
  \bibinfo{author}{F.~M.~A. Pires}, \bibinfo{author}{S.~T. Pinho},
\newblock \bibinfo{title}{Micromechanical analysis of polymer composites
  reinforced by unidirectional fibres: Part i-constitutive modelling},
\newblock \bibinfo{journal}{International Journal of Solids and Structures}
  \bibinfo{volume}{50} (\bibinfo{year}{2013}) \bibinfo{pages}{1897--1905}.
\bibitem[{McKay et~al.(1979)McKay, Beckman, and Conover}]{McKay1979}
\bibinfo{author}{M.~D. McKay}, \bibinfo{author}{R.~J. Beckman},
  \bibinfo{author}{W.~J. Conover},
\newblock \bibinfo{title}{A comparison of three methods for selecting values of
  input variables in the analysis of output from a computer code},
\newblock \bibinfo{journal}{Technometrics} \bibinfo{volume}{21}
  (\bibinfo{year}{1979}) \bibinfo{pages}{239}. \URLprefix
  \url{https://www.jstor.org/stable/1268522?origin=crossref}.

\end{thebibliography}

\end{document}